\documentclass[twocolumn]{aastex63}
\usepackage{diagbox}
\usepackage{multirow}
\usepackage{amsmath}
\usepackage{threeparttable}
\usepackage{color}
\usepackage{array}
\usepackage[T1]{fontenc}

\begin{document}

\title{Reconstruction of X-Ray Afterglow Light Curves of GRBs and its implication for constraining Cosmological Parameters}

\author{Yu-Qi Zhou}
\author[0000-0003-0672-5646]{Shuang-Xi Yi$^{\dag}$}
\author{Yu-Peng Yang}
\author{Jia-Lun Li}
\affiliation{School of Physics and Physical Engineering, Qufu Normal University, Qufu 273165, China; yisx2015@qfnu.edu.cn,ypyang@qfnu.edu.cn}
\author{Jian-Ping Hu}
\affiliation{School of Physics, Xi'an Jiaotong University, Xi'an 710049, China}
\author{Yan-Kun Qu}
\affiliation{School of Physics and Physical Engineering, Qufu Normal University, Qufu 273165, China; yisx2015@qfnu.edu.cn,ypyang@qfnu.edu.cn}
\author{Fa-Yin Wang}
\affiliation{School of Astronomy and Space Science, Nanjing University, Nanjing 210023, China£» fayinwang@nju.edu.cn}

\begin{abstract}

Gamma-ray bursts (GRBs) serve as important cosmological probes, whose X-ray afterglow light curves (LCs) may exhibit a plateau phase (with temporal slope $\alpha$ between 0 and 0.5) that may originate from magnetar energy injection. Similar to Type Ia Supernovae, GRBs with a common physical origin can be used as standardizable candles for cosmological studies. However, observational gaps in GRB light curves introduce significant uncertainties in plateau parameter estimation, thereby affecting cosmological constraints. In this work, we employ a stochastic reconstruction technique to reconstruct the X-ray afterglow LCs for 35 GRB samples exhibiting plateau features, generating 50 simulated data points for each LC. Using the reconstructed LCs, we calibrate three luminosity correlations: the $L_0$-$t_b$, $L_0$-$t_b$-$E_{p,i}$, and $L_0$-$t_b$-$E_{\gamma,\mathrm{iso}}$ relation, which are then applied to constrain both flat and non-flat $\Lambda$CDM cosmological models. The main results include: (i) the $L_0$-$t_b$ relation yields a slope $b \approx -1$, supporting a constant magnetar energy injection rate; (ii) light curve reconstruction has limited impact on cosmological parameter constraints; (iii) for the flat $\Lambda$CDM model constrained by the $L_0$-$t_b$-$E_{p,i}$ relation, the precision of $\Omega_m$ improves by 6.25\%; For the non-flat $\Lambda$CDM model constrained by the $L_0$-$t_b$-$E_{p,i}$ relation, the precision of $\Omega_\Lambda$ improves by 1.01\%. Our findings suggest that increasing the number of LC data points provides limited improvement to cosmological constraints, while expanding the sample size of GRBs with identical physical origins may be more crucial.

\end{abstract}

\keywords{Gamma-ray bursts; Cosmology; Magnetar}

\section{Introduction}
\label{section:1}

Gamma-ray bursts (GRBs) are among the most extreme and energetic events in the universe, providing crucial information about the early universe, galaxy evolution, and the formation of black holes and neutron stars.  The detection of GRBs at redshifts as high as $z$=8.2 (\citealp{2009Natur.461.1258S}; \citealp{2009Natur.461.1254T}) and even $z$=9.4 (\citealp{2011ApJ...736....7C}), significantly exceeding the maximum redshift observed for Type Ia supernovae (SNe Ia, around $z\sim2$; \citealp{2018ApJ...859..101S, 2022ApJ...938..113S}), establishes them as exceptional probes of the early universe. GRBs are typically classified into long GRBs (LGRBs, $T_{90} >$ 2s) and short GRBs (SGRBs, $T_{90} <$ 2s) (\citealp{1993ApJ...413L.101K}; \citealp{2013ApJ...763...15Q}), generally attributed to the collapse of massive stars and the mergers of compact star binaries, respectively. Alternatively, they can be categorized into Type-I and Type-II based on their progenitors (\citealp{2007ApJ...655L..25Z}).

GRBs exhibit two distinct emission phases: the prompt phase and the afterglow phase. The prompt phase, characterized by its intense gamma-ray emission, typically lasts from milliseconds to several seconds, although in long GRBs (LGRBs), durations can extend to minutes. The afterglow phase follows, spanning hours to weeks or even longer, and is observable across X-ray, optical, and radio wavelengths.

The afterglow phase, observable across various wavelengths, exhibits complex behavior. This complexity is highlighted by the presence of multiple phases in the X-ray afterglow light curves (LCs)(\citealp{2006ApJ...642..354Z}; \citealp{2006ApJ...642..389N}). One prominent feature is the extended plateau phase, often observed following the prompt emission. The presence and properties of this plateau are intimately linked to the nature of the GRB central engine, the power source driving the relativistic outflow.

Several models have been proposed to explain the origin of the plateau, including those invoking long-lived activity from a black hole accretion disk and models invoking internal shocks within the jet (\citealp{2003ApJ...592.1042I}; \citealp{2002ApJ...577..311K}; \citealp{2003qftn.book.....Z}; \citealp{2021ApJ...908..242D}; \citealp{2021MNRAS.507.1047Y}). However, these models often struggle to account for the duration and flares of some observed plateaus (\citealp{2016ApJS..224...20Y,2022ApJ...924...69Y}). In contrast, the millisecond magnetar model offers a compelling alternative. This model postulates that a newly formed, rapidly rotating magnetar, endowed with a strong magnetic field, provides sustained energy injection through its spin-down, naturally explaining the extended plateau phase (\citealp{1992Natur.357..472U}; \citealp{2001ApJ...552L..35Z}). The magnetar's gradual loss of rotational energy powers the observed luminosity, with the duration of the plateau determined by the initial spin period and magnetic field strength of the magnetar. This model, unlike some alternatives, directly links the observed properties of the plateau to the physical characteristics of the central engine, making it particularly attractive for its predictive power and its ability to explain diverse plateau morphologies. Understanding the physical mechanisms responsible for these plateaus, and accurately characterizing their properties, is crucial for utilizing GRBs as cosmological probes.

It is frequently observed that the X-ray afterglow LCs of GRBs exhibit numerous gaps. These gaps can appear at the start of the plateau phase, during the plateau phase, or at the end of the plateau phase. When a GRB occurs, gaps can appear in the GRB light curve if the satellite's orbital position is unfavorable, or if the GRB is obscured by the Earth, preventing the satellite from receiving GRB photons. Furthermore, insufficient rapid response observations, atmospheric conditions, and potential failures or inaccuracies of the detection instruments can also lead to gaps in the X-ray afterglow LCs of GRBs.

The existence of these gaps affects the precise determination of the plateau parameters. Light curve reconstruction (LCR) method have been developed to mitigate this issue. The reconstruction approaches and Machine Learning technology have become valuable tools in multiple areas of astronomy (\citealp{2022Wang}; \citealp{2023arXiv231002602S}; \citealp{2023MNRAS.525.5204B}; \citealp{2024ApJS..271...22D}; \citealp{2024MNRAS.534...56Z}; \citealp{2024arXiv240115632C}; \citealp{2025A&A...698A..92N}; \citealp{2025ApJS..276...62C}). \cite{2023ApJS..267...42D} showed that after filling the gaps in the LCs of GRBs, the uncertainties of the afterglow plateau parameters generally decreased to varying degrees. It is noteworthy that \cite{2022MNRAS.514.1828D} demonstrated that increasing the sample size and reducing the uncertainty of the plateau parameters allows for achieving the same precision in $\Omega_m$ as would be expected in the next decade or more. This demonstrates the great attraction of reconstructing GRB afterglow LCs for cosmological research.

Similar to the SNe Ia used as standard candles for cosmological studies (\citealp{1998AJ....116.1009R}; \citealp{1999ApJ...517..565P}; \citealp{2018ApJ...859..101S}), the luminosity correlations of GRBs (\citealp{2002A&A...390...81A}; \citealp{2004ApJ...616..331G}; \citealp{2004ApJ...609..935Y}; \citealp{2005ApJ...633..611L}; \citealp{2008MNRAS.391L..79D, 2016ApJ...825L..20D}; \citealp{2015A&A...582A.115I}; \citealp{2011ApJ...730..135D}; \citealp{2020ApJ...904...97D}; \citealp{2022MNRAS.516.1386C}; \citealp{2022ApJS..261...25D}) can be used as a cosmological tool to investigate magnetar energy injection and to constrain cosmological models (\citealp{2004ApJ...613L..13G}; \citealp{2006MNRAS.369L..37L}; \citealp{2015NewAR..67....1W}; \citealp{2013MNRAS.436...82D}; \citealp{2023ApJ...951...63D}; \citealp{2024JCAP...08..015A}; \citealp{2025NewAR.10001712B}). In earlier studies, luminosity correlations were calibrated by assuming a specific cosmological model (\citealp{2004ApJ...612L.101D}). However, using these correlations to subsequently constrain cosmological models introduces a circularity problem. Several methods have been proposed to address this issue (\citealp{2008AIPC.1065..367L}; \citealp{2008MNRAS.391..577A}; \citealp{2008PhRvD..78l3532W}; \citealp{2024arXiv241103773M}). The Gaussian Process (GP) method is a widely applied approach for resolving the circularity problem, which is a data smoothing technique based on Bayesian statistics (\citealp{2012PhRvD..86h3001S}, \citeyear{2012JCAP...06..036S}; \citealp{2013arXiv1311.6678S}; \citealp{2014MNRAS.441L..11B}; \citealp{2018MNRAS.474..313L}).

Therefore, we employ the stochastic reconstruction method mentioned in \cite{2023ApJS..267...42D} to reconstruct the LCs of a selected sample of GRB X-ray afterglows that exhibit plateau features. Following this reconstruction, we process to calibrate the $L_{\rm 0}-t_{\rm b}$ correlation, the $L_{\rm 0}-t_{\rm b}-E_{\rm p,i}$ correlation, and the $L_{\rm 0}-t_{\rm b}-E_{\rm \gamma,iso}$ correlation. These calibrated correlations are then utilized to constrain the flat $\Lambda$CDM model and the non-flat $\Lambda$CDM model.

The structure of this paper is as follows. The next section presents the methods we used. In Sect. \ref{section:3}, we introduce the correlations used in this paper and calibrate them. In Sect. \ref{section:4} we place constraints on cosmological parameters. Our results are presented in Sect. \ref{section:5}. Finally, a summary of our work is presented in Sect. \ref{section:6}

\section{Method}
\label{section:2}

Our aim is to fill the gaps in the GRB X-ray LCs. To ensure the generated data points resemble actual observations, we introduce a noise function that allows the filled data points to fluctuate around the fitted line. Here, we assume that the occurrence of all plateaus is related to the re-injection of energy from magnetars, meaning that all the X-ray LCs discussed in this paper are fitted using the magnetar model.

\subsection{The GRB samples and the magnetar model}

\cite{2022ApJ...924...97W} explored the possibility of using long GRBs with plateau features as distance indicator to study cosmology. They found 10 long GRB samples that fit the magnetar model very well and defined them as Gold Samples. There are 21 cases where the data do not show clear plateau characteristics, but there may have been a plateau combining with the BAT data, and this group is referred to as the Silver Sample.

It is worth noting that these samples are as of June 2020 data. Swift has observed more GRBs, therefore, we have added four more samples (GRB 201221A, GRB 210210A, GRB 210610B, GRB 241030A), supplementing the plateau samples that are consistent with the magnetar model. We present some information about these four GRBs in Table \ref{tab:addlabel}. In this analysis, the best-fit values of $F_0$ and $t_b$ are determined utilizing the Monte Carlo Markov Chain (MCMC) method, implemented through the publicly available \texttt{emcee} package (\citealp{2013PASP..125..306F}). Note that all the fittings in this work are performed using this method. The flux ($F_0$) and break time ($t_b$) in Table \ref{tab:addlabel} are obtained from fitting the magnetar model. In this model, the luminosity evolving with time can be written as (\citealp{1998A&A...333L..87D} )
\begin{equation}\label{}
L=L_{0}\times \frac{1}{\left(1+t/t_{b}\right)^2 } \simeq
\begin{cases}
L_{0} &, {t\ll   t_{b}}\\
L_{0} \left ( t/t_{b} \right )^{-2}&, t\gg   t_{b}
\end{cases}
\label{eq:1}
\end{equation}
where $L_0$ and $t_b$ represent the characteristic spin-down luminosity and end time of the plateau, respectively.

\subsection{The reconstruction method with the functional form}

Three principal methodologies dominate GRB LC reconstruction: (i) conventional non-parametric techniques (such as Gaussian Processes), (ii) physics-based parametric models (including the W07 model, the BPL model, and the magnetar model), and (iii) machine learning approaches incorporating physical constraints (exemplified by neural networks and symbolic regression algorithms)(\citealp{2024arXiv241220091M}). Compared with others, the magnetar model is based on a possible physical mechanism during the plateau period. In this work, we assume that each GRB conforms to the magnetar model, which implies that the luminosity $L$ and the time $t$ of each GRB have a relationship given by Equation (\hyperref[eq:1]{1}).

The flux residual for each LC can be computed as
\begin{equation}\label{eq:2}
\log_{10}F_t^{res}=\log_{10}F_{t}^{obs} -\log_{10}f\left ( t \right )
\end{equation}
where $\log_{10}F_t^{res}$ is the logarithm of the flux residual, $\log_{10}f\left ( t \right )$ is the magnetar model flux at time $t$, and $\log_{10}F_{t}^{obs}$ is the observed log flux at time $t$.

Using this formula, we can obtain the corresponding histogram of the flux residual distribution for the GRBs studied by us. We adopt that the flux residuals follow a Gaussian distribution,
\begin{equation}\label{}
P\left ( x  \right ) =\frac{1}{\sigma \sqrt{2\pi } }  e^{-\left ( x-\mu  \right )^2/2\sigma  }
\label{eq:3}
\end{equation}
where $\mu$ is the mean of this distribution, $\sigma$ is the standard deviation. Since the flux residuals of the original data points follow a Gaussian distribution, the flux residuals of the reconstructed data points should also follow a Gaussian distribution. We fit the GRBs LCs with a Gaussian distribution to obtain the value of $\mu$ and $\sigma$ corresponding to each GRB, and further generate reconstruction data points with the help of these values.

For a given time $t$, the reconstructed flux is related to the flux residual Gaussian distribution and the noise, namely,
\begin{equation}\label{}
\log{F^{res}_t}=\log f\left ( t \right ) +\left (  1+n\right ) \times RV_{\mathcal{N}}
\label{eq:4}
\end{equation}
where, $\log{F^{res}_t}$ is the reconstructed flux at time $t$, $\log f\left ( t \right )$ is the magnetar model flux at time $t$, $n$ is the noise level, and $RV_{\mathcal{N}}$ is a random value from the normal distribution. Here, we set $n=10\%$ for our analysis.

We generated 50 reconstructed data points based on the fitted curve. \cite{2023ApJS..267...42D} reconstructed 100 data points for the X-ray LCs of GRBs. However, at this stage, the data points in the LCs are distributed too densely, potentially exceeding the detector's sensitivity threshold (\citealp{2004ApJ...611.1005G}; \citealp{2004SPIE.5165..175B}; \citealp{2005SSRv..120..165B}). In general, as the afterglow progresses to later stages, the number of photons detected by the instrument decreases. To ensure that the reconstruction aligns more closely with the actual physical scenario, we need to avoid an excessively large number of reconstructed data points. At the same time, too few data points would inadequately fill the gaps in the LCs. Therefore, we select an intermediate value of 50 reconstructed data points as the optimal balance. We then use the $geomspace$ function in Python, setting the minimum and maximum values to the beginning of the plateau and the end of the afterglow, respectively. The images of the 35 reconstructed GRBs are presented in Figure $\ref{fig:1}$.

\begin{figure*}[tp]
\centering

\resizebox{45mm}{!}{\includegraphics[]{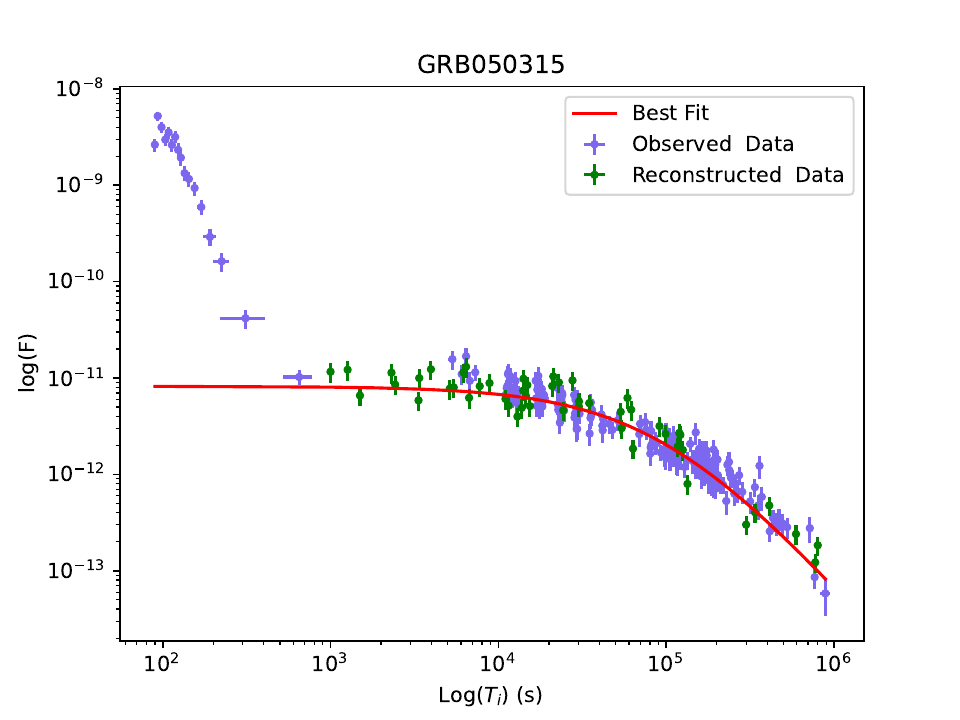}}%
\resizebox{45mm}{!}{\includegraphics[]{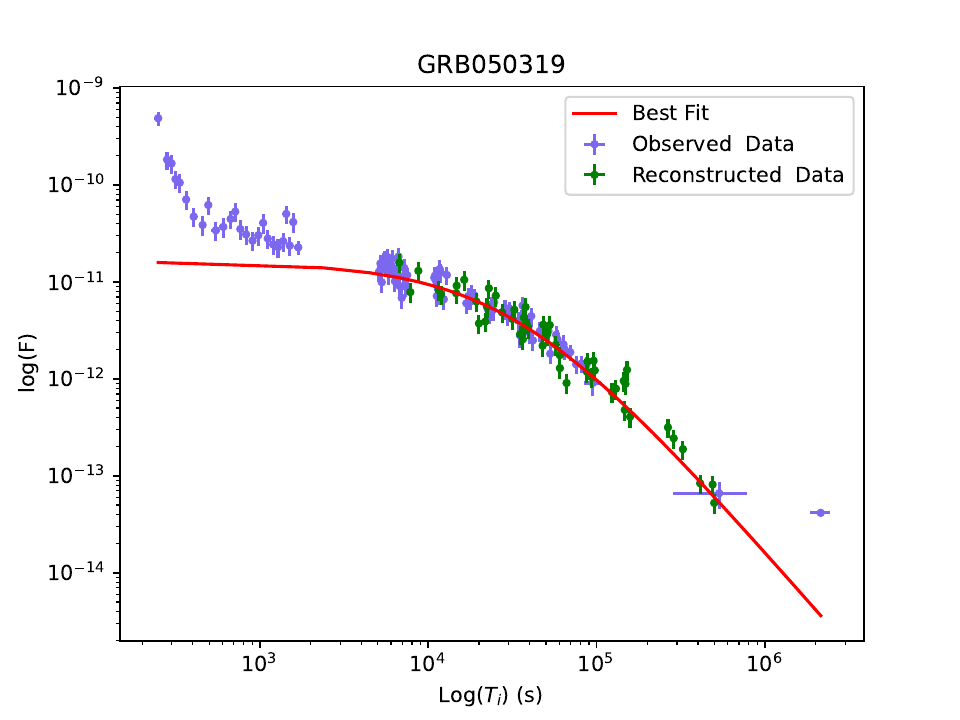}}%
\resizebox{45mm}{!}{\includegraphics[]{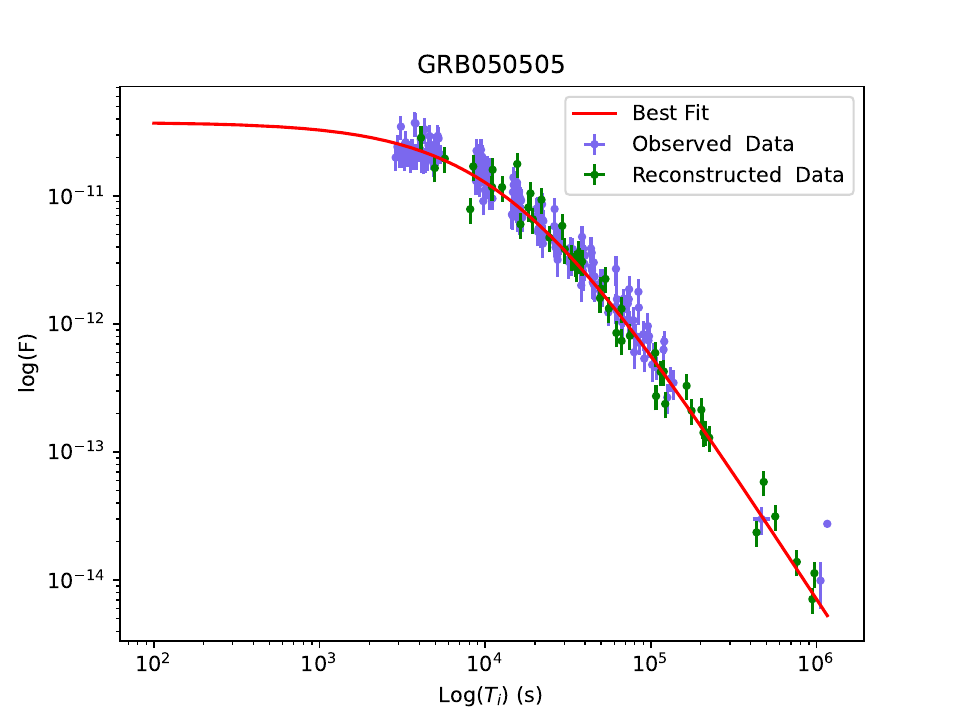}}%
\resizebox{45mm}{!}{\includegraphics[]{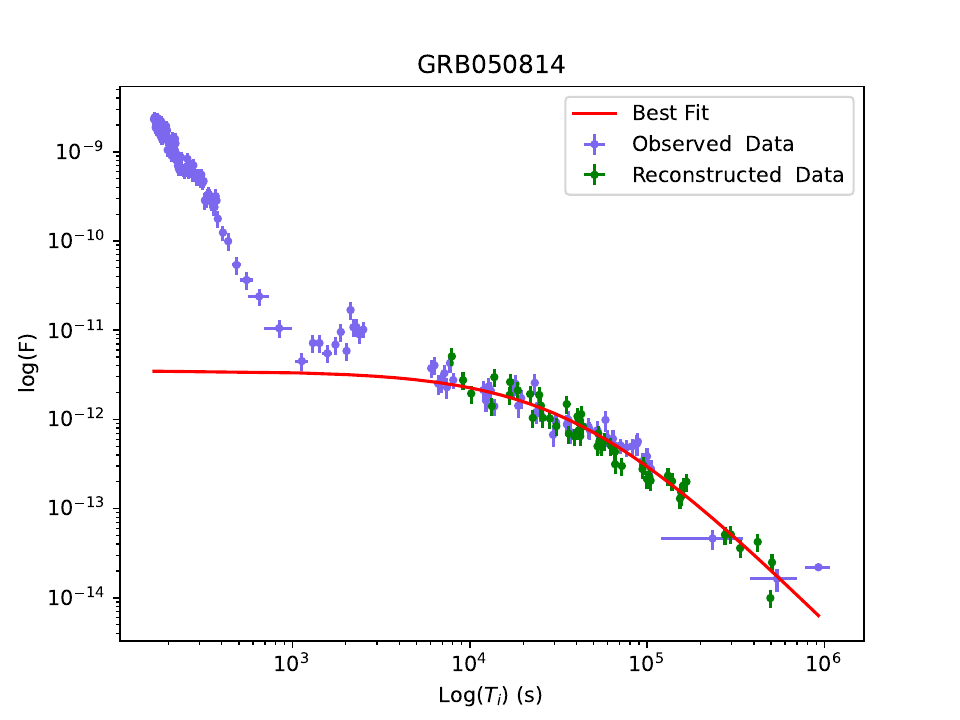}}\\

\resizebox{45mm}{!}{\includegraphics[]{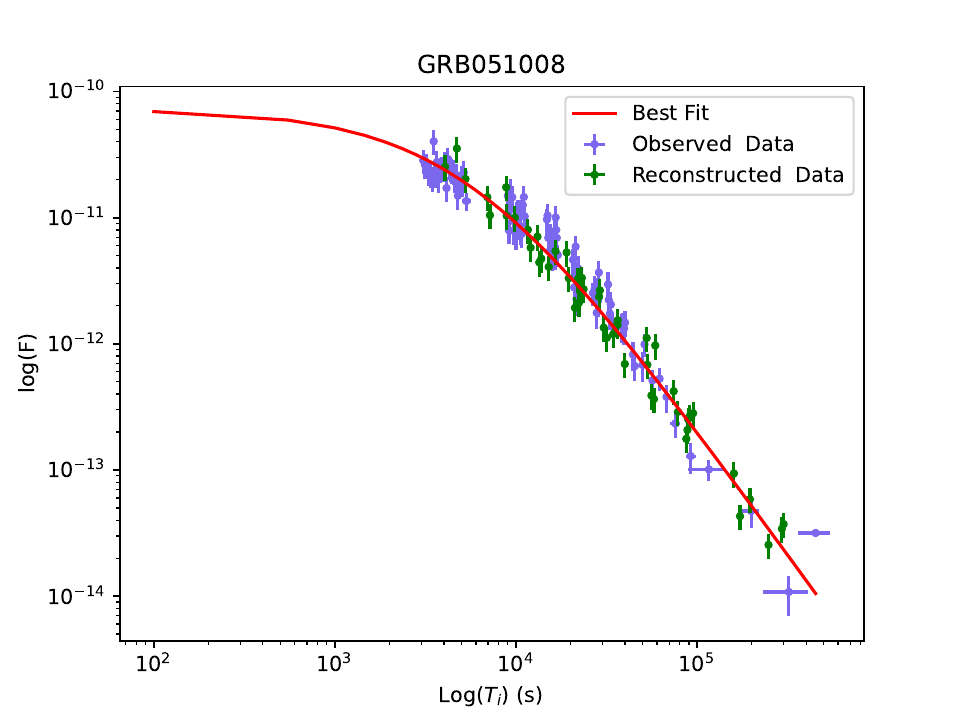}}%
\resizebox{45mm}{!}{\includegraphics[]{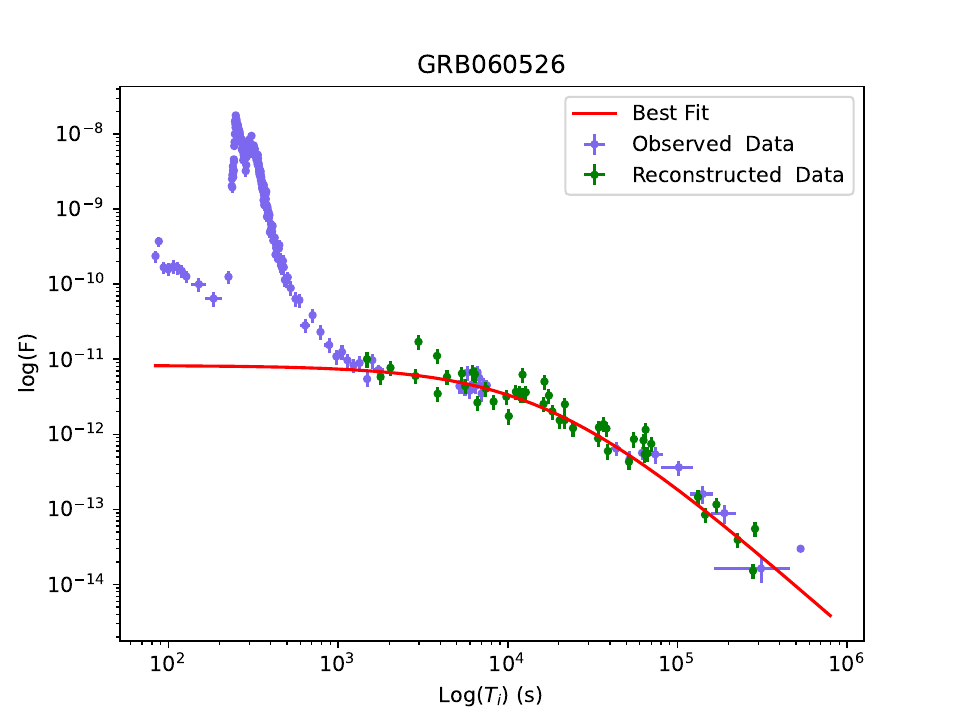}}%
\resizebox{45mm}{!}{\includegraphics[]{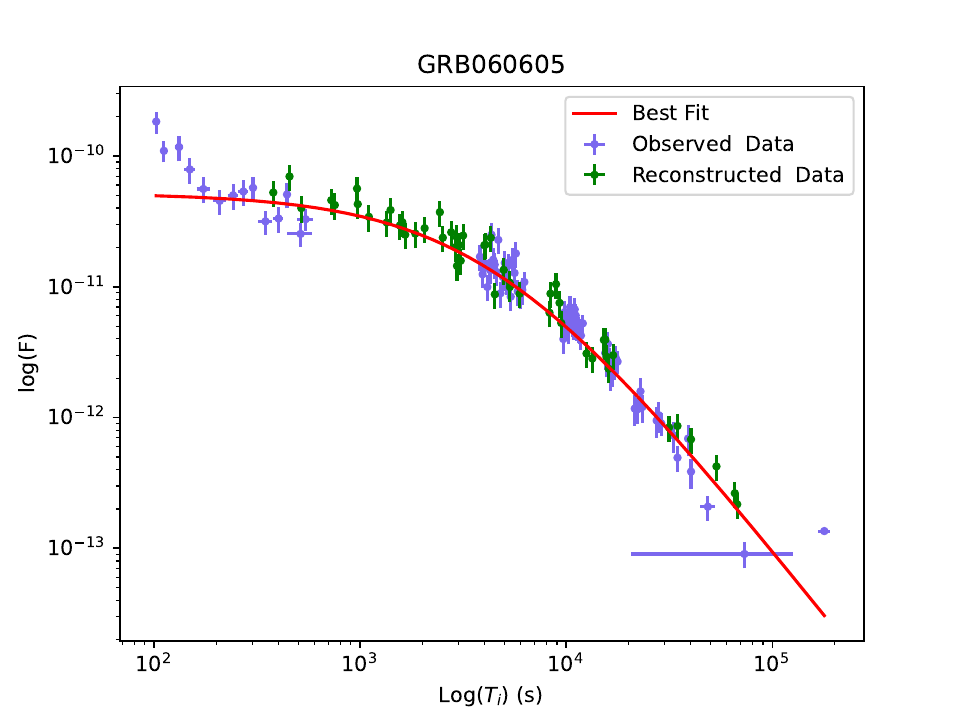}}%
\resizebox{45mm}{!}{\includegraphics[]{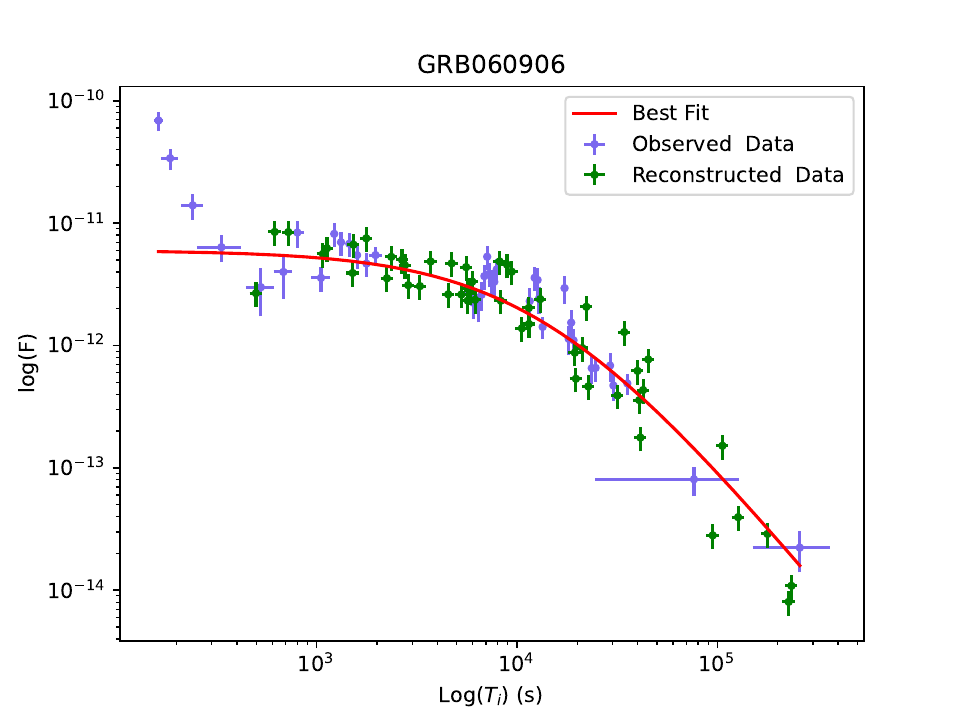}}\\

\resizebox{45mm}{!}{\includegraphics[]{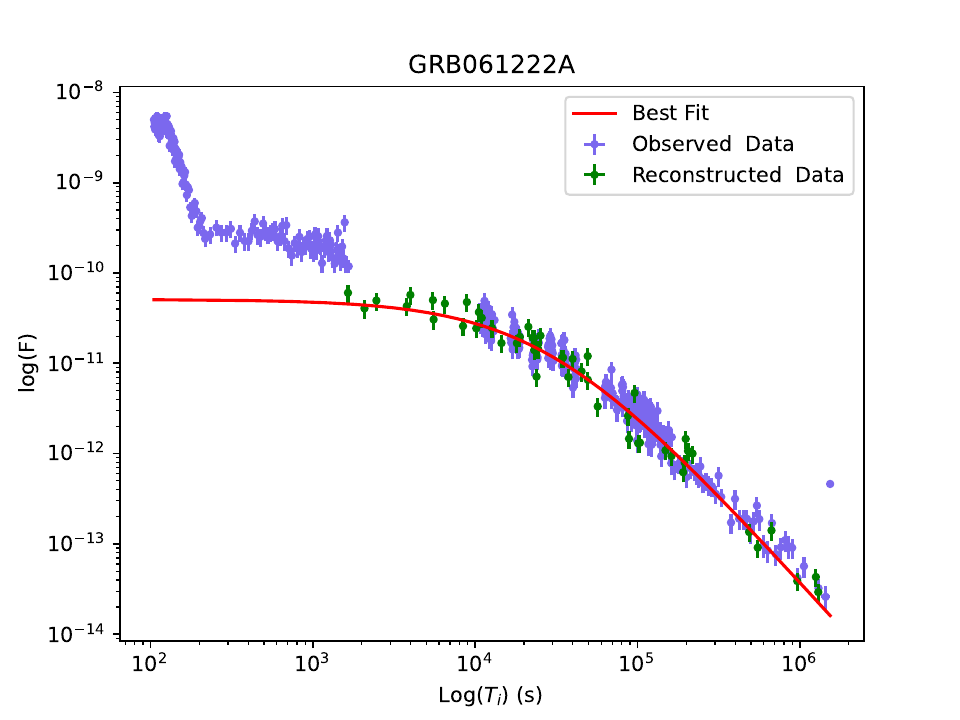}}%
\resizebox{45mm}{!}{\includegraphics[]{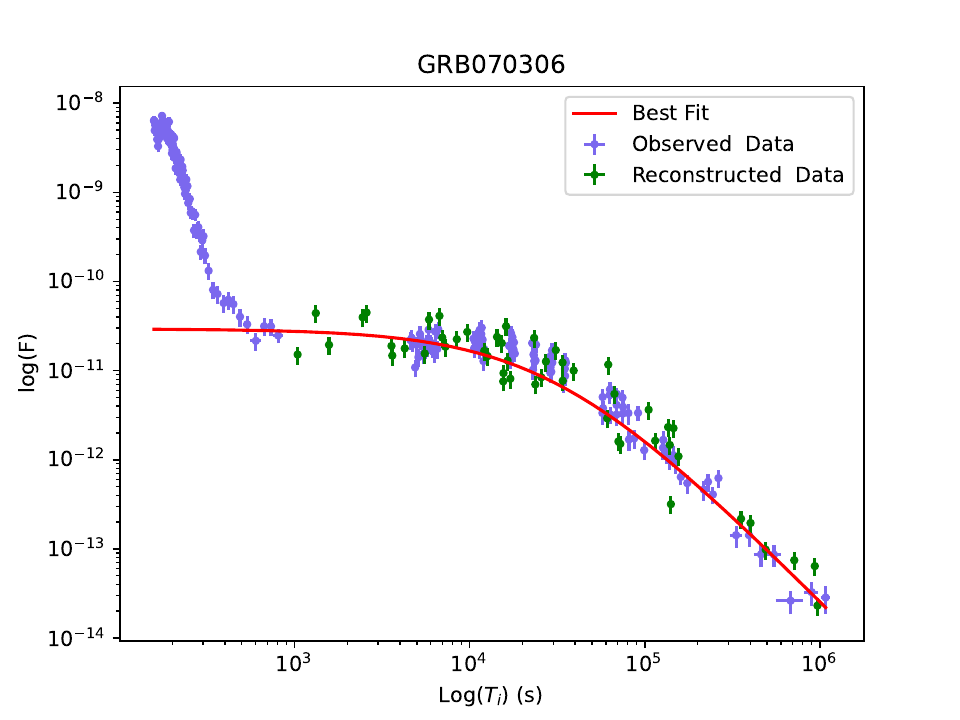}}%
\resizebox{45mm}{!}{\includegraphics[]{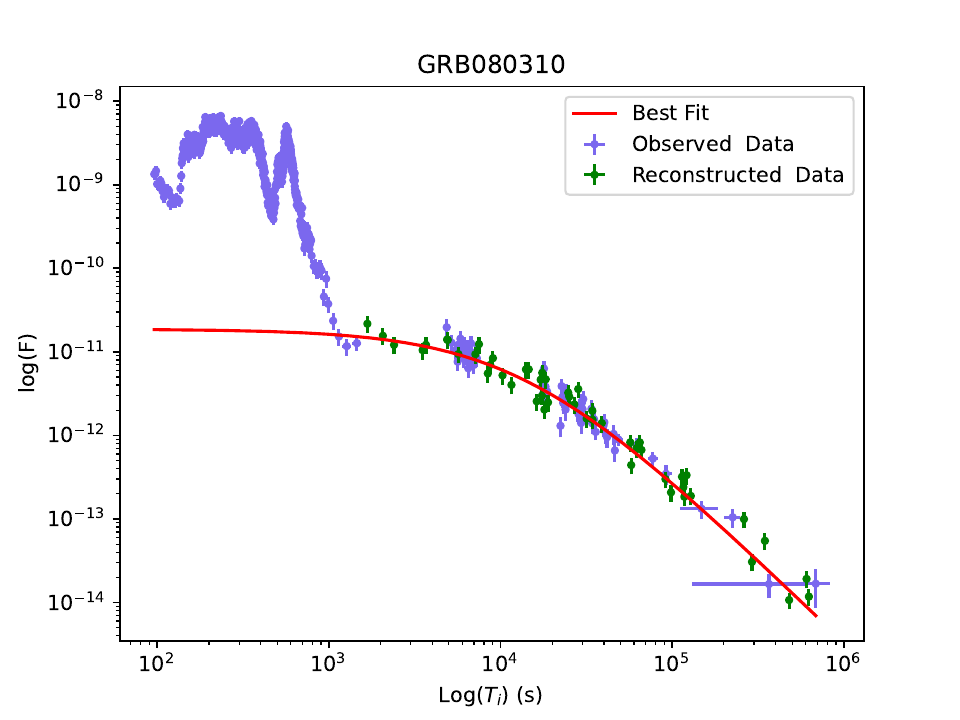}}%
\resizebox{45mm}{!}{\includegraphics[]{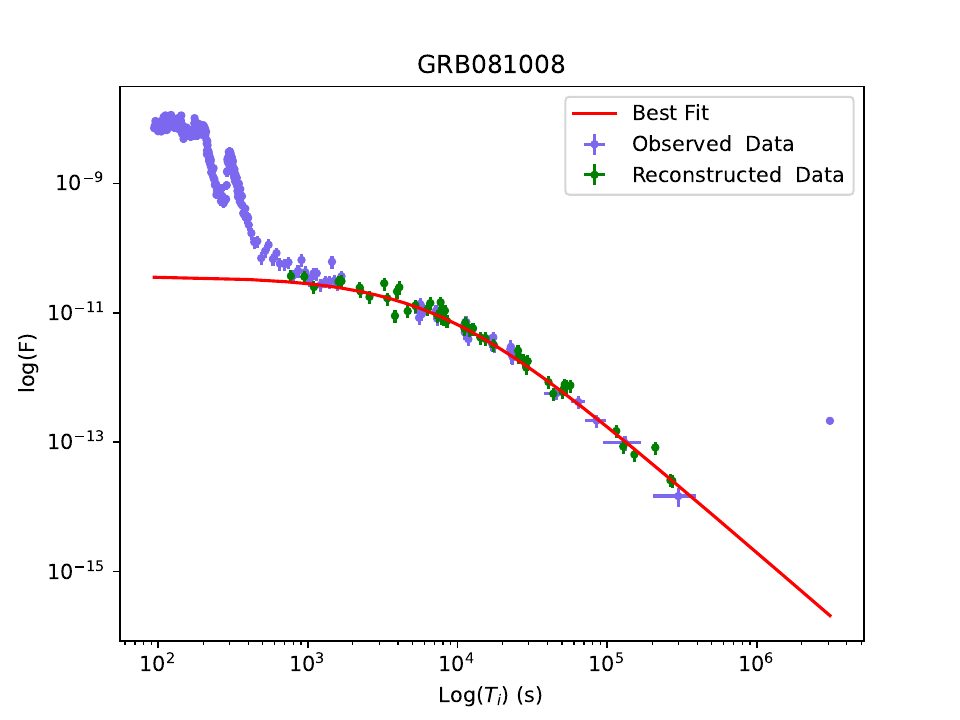}}\\

\resizebox{45mm}{!}{\includegraphics[]{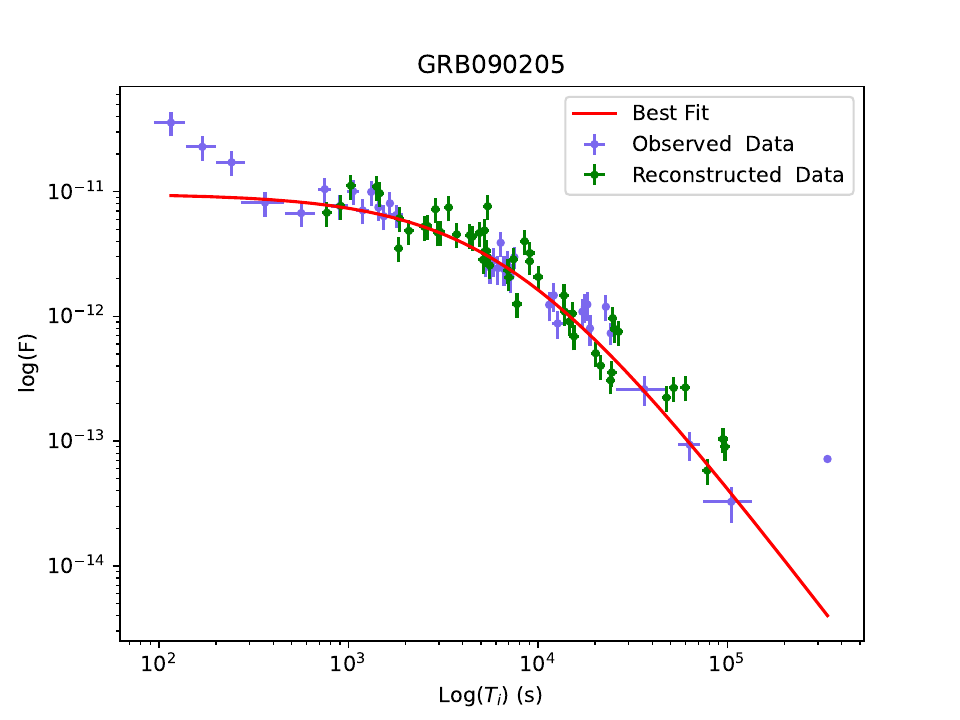}}%
\resizebox{45mm}{!}{\includegraphics[]{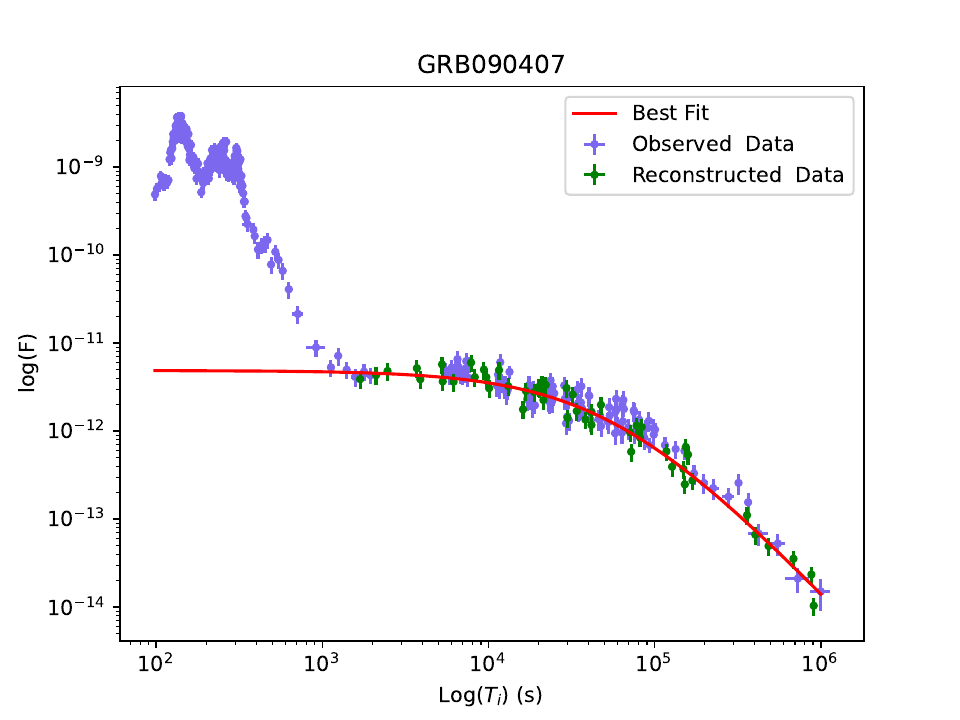}}%
\resizebox{45mm}{!}{\includegraphics[]{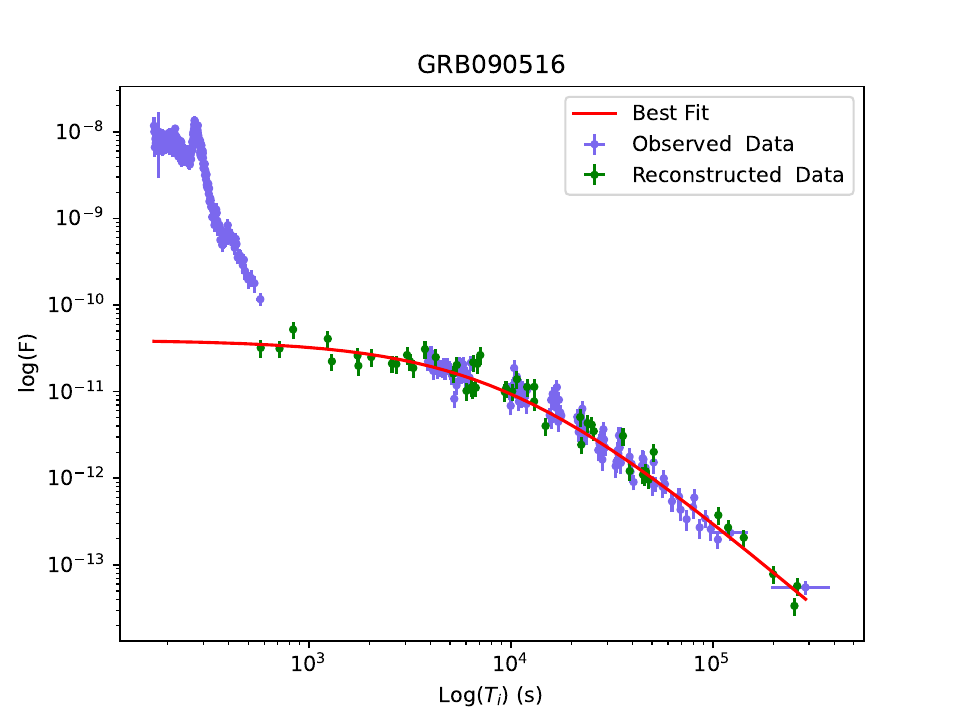}}%
\resizebox{45mm}{!}{\includegraphics[]{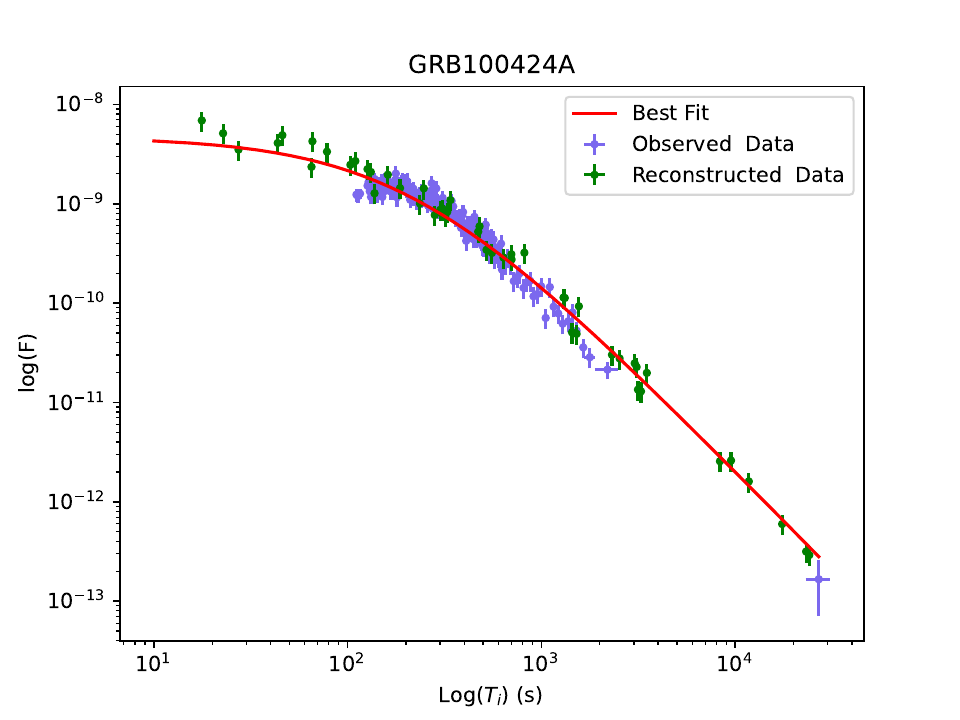}}\\

\resizebox{45mm}{!}{\includegraphics[]{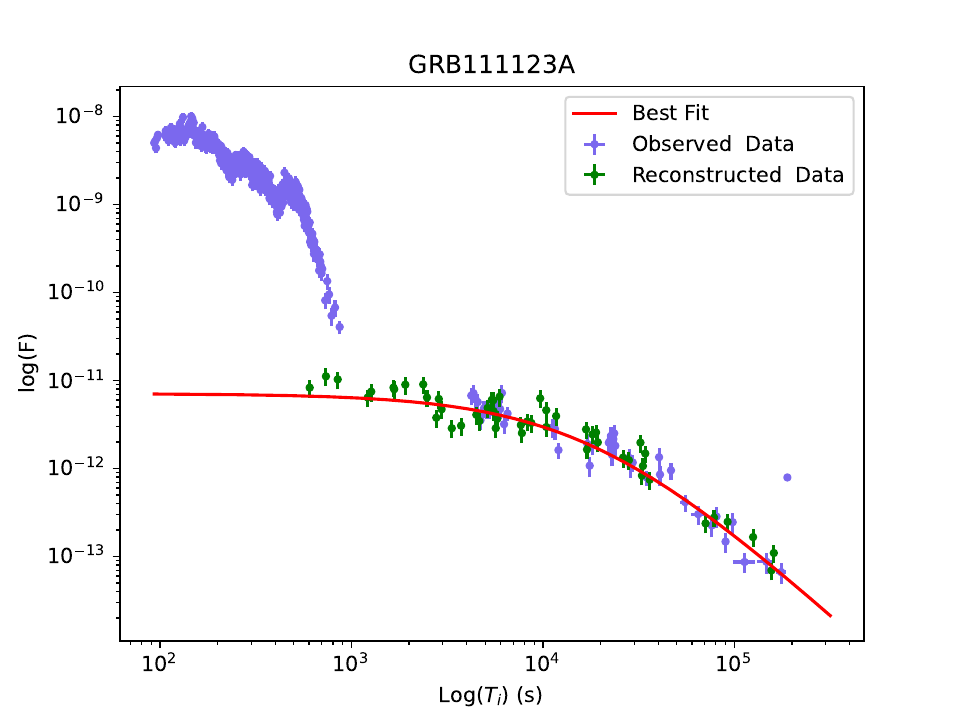}}%
\resizebox{45mm}{!}{\includegraphics[]{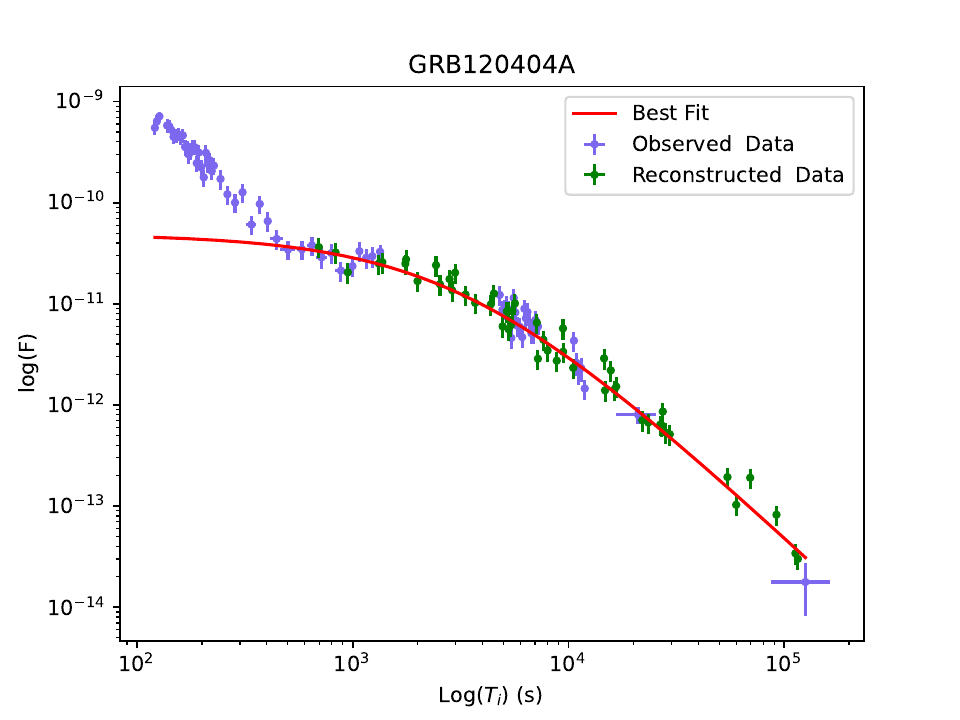}}%
\resizebox{45mm}{!}{\includegraphics[]{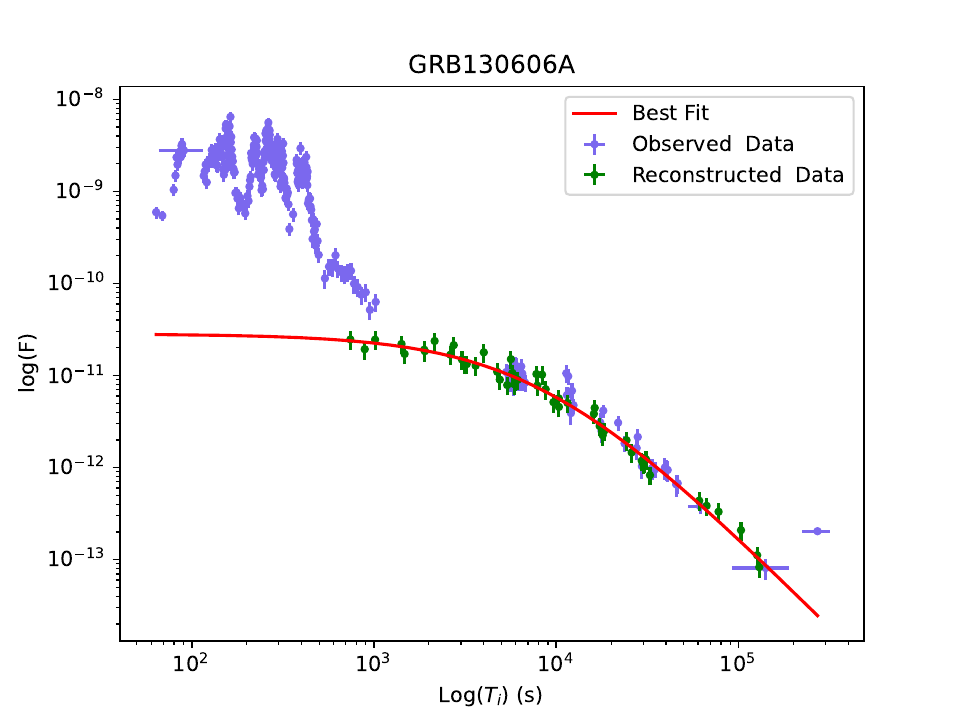}}%
\resizebox{45mm}{!}{\includegraphics[]{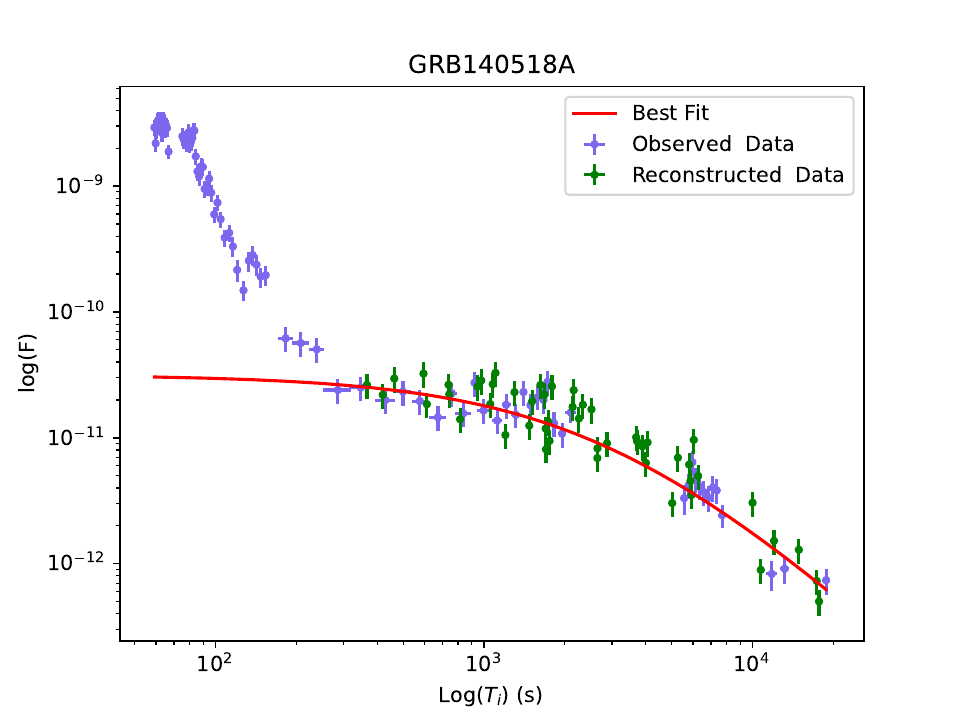}}\\

\resizebox{45mm}{!}{\includegraphics[]{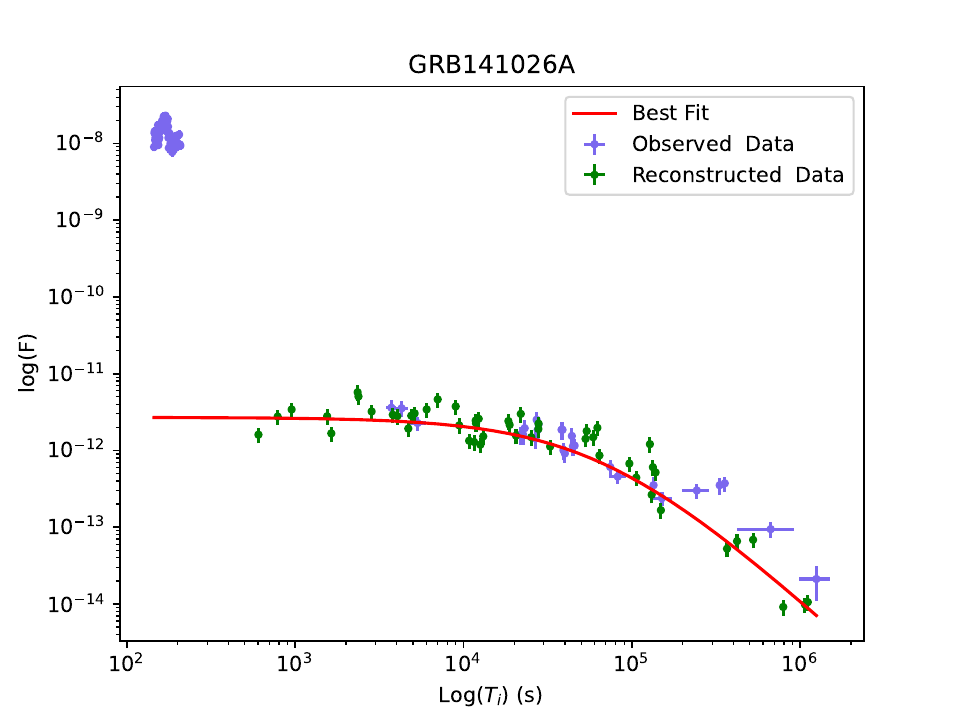}}%
\resizebox{45mm}{!}{\includegraphics[]{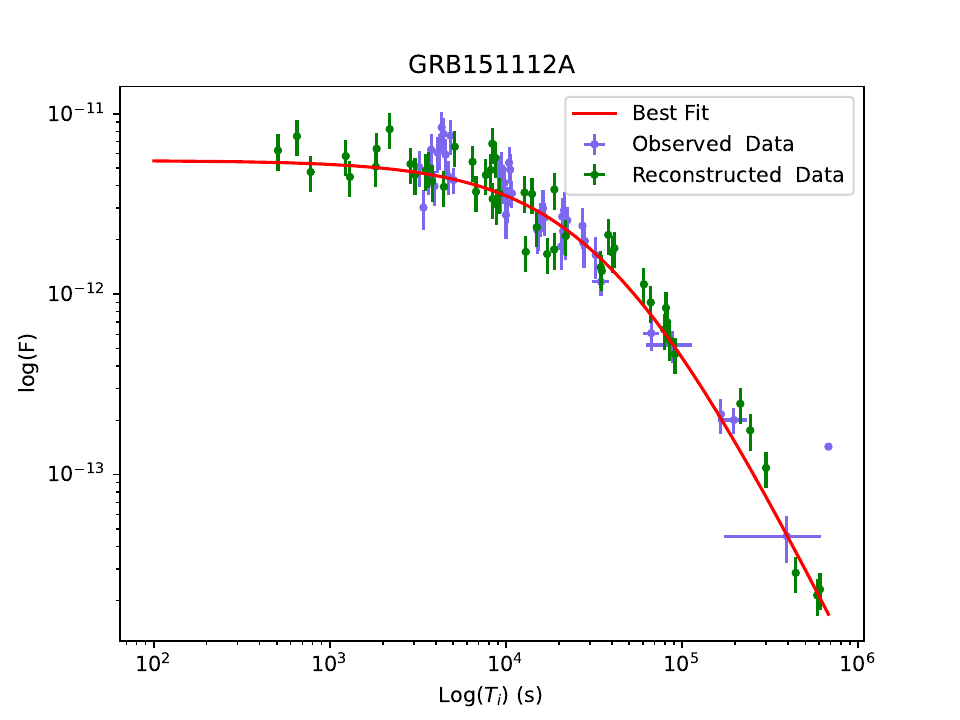}}%
\resizebox{45mm}{!}{\includegraphics[]{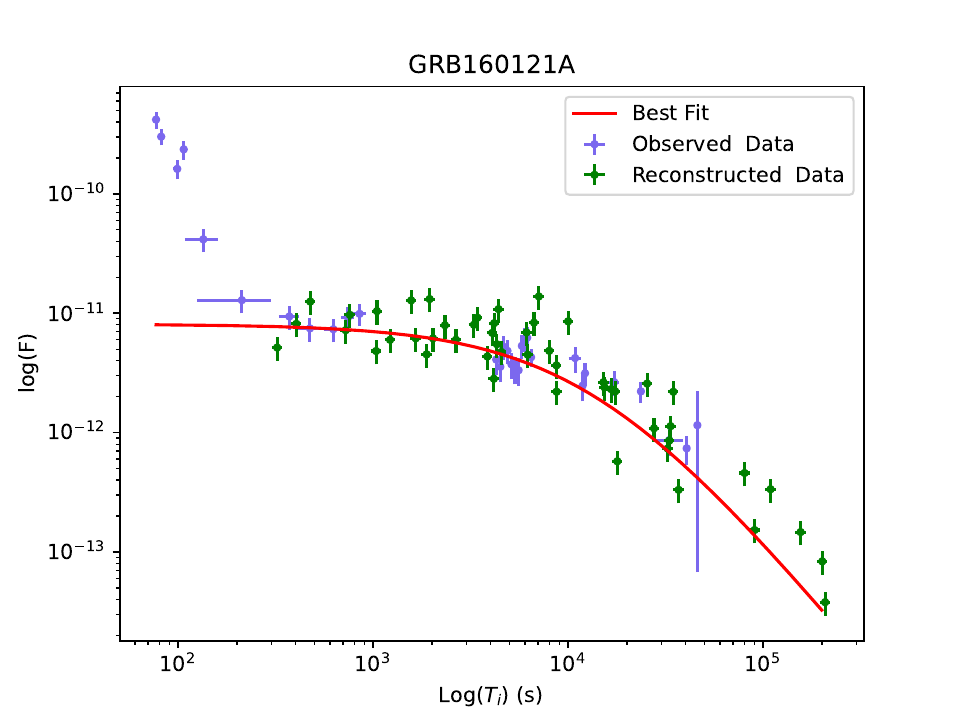}}%
\resizebox{45mm}{!}{\includegraphics[]{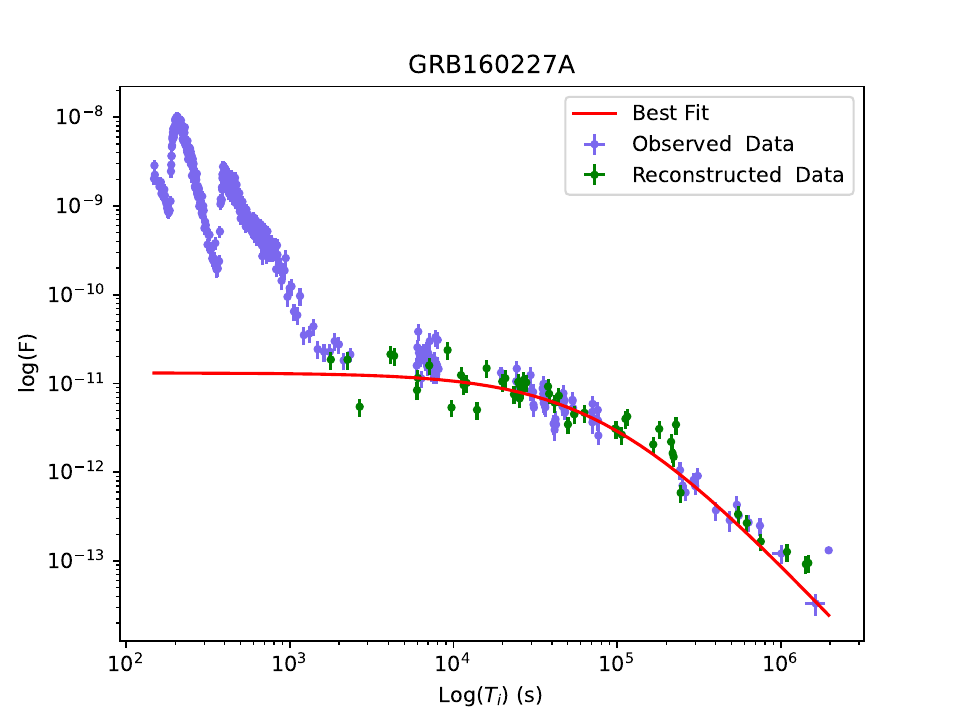}}\\

\resizebox{45mm}{!}{\includegraphics[]{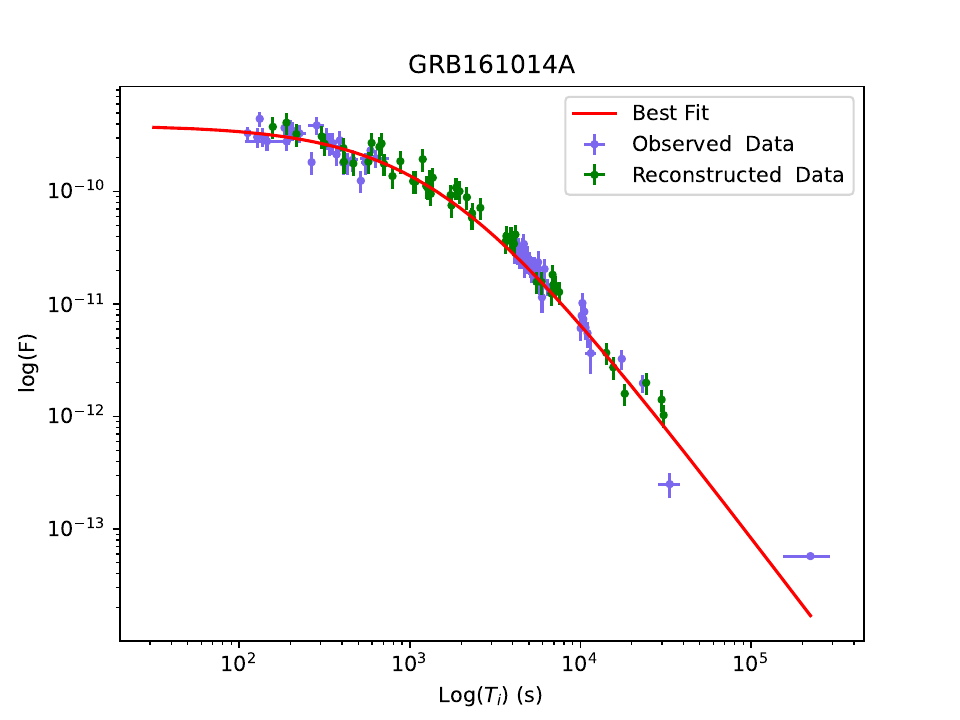}}%
\resizebox{45mm}{!}{\includegraphics[]{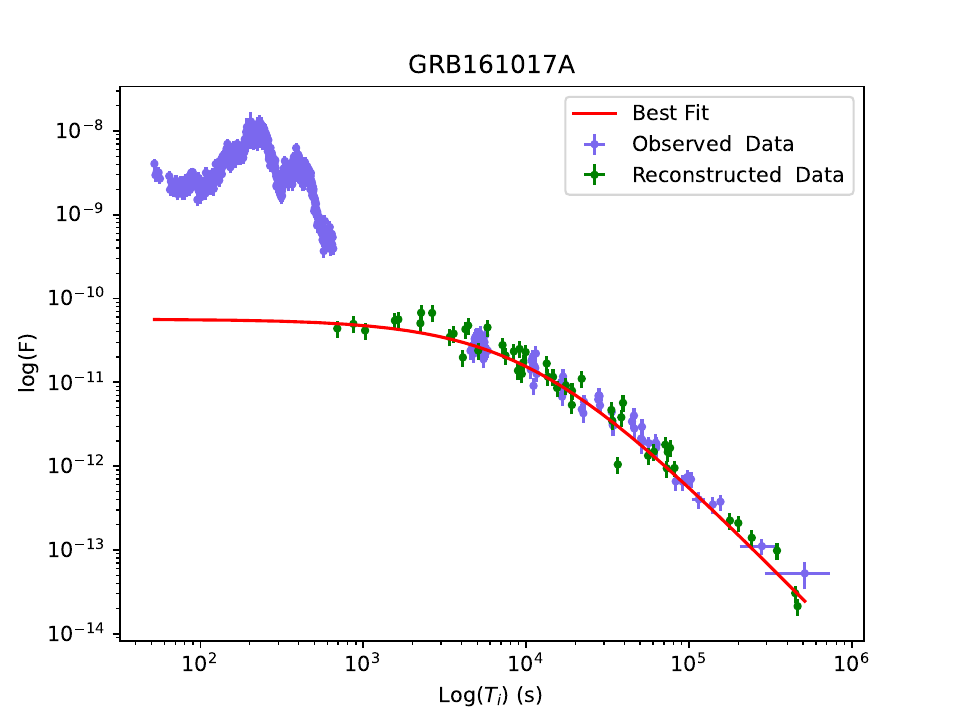}}%
\resizebox{45mm}{!}{\includegraphics[]{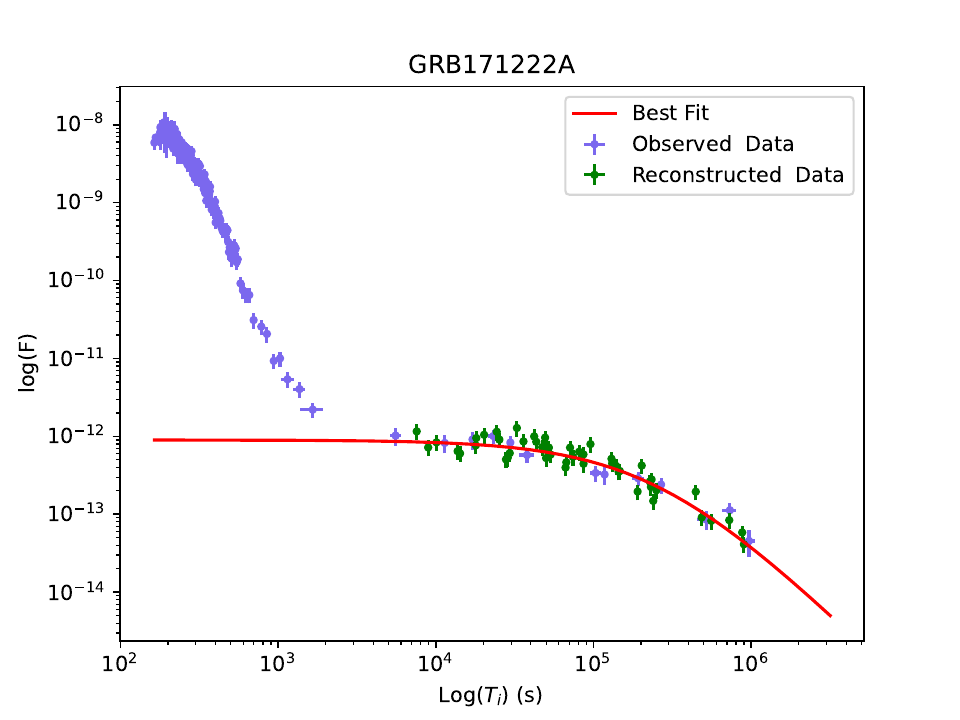}}%
\resizebox{45mm}{!}{\includegraphics[]{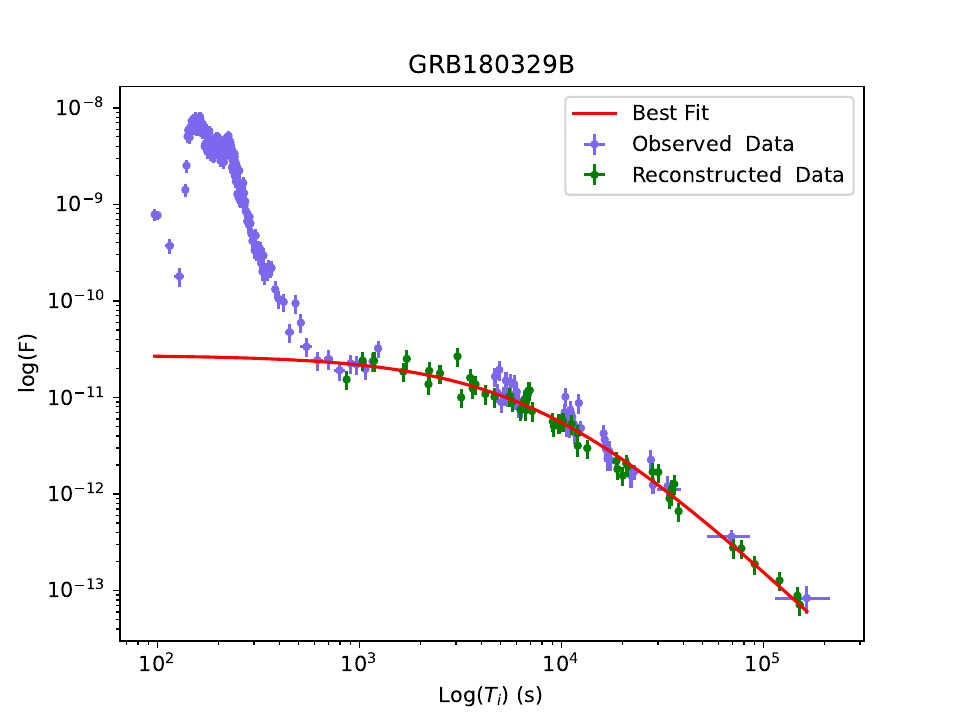}}\\
\caption{The X-ray LCs of 35 GRB samples after reconstruction. The blue points represent the original observational data, the green points represent the reconstructed data (50 points), and the red solid lines are the best fit using the magnetar model.}
\label{fig:1}
\end{figure*}

\addtocounter{figure}{-1}

\begin{figure*}[ht!]
\centering

\resizebox{45mm}{!}{\includegraphics[]{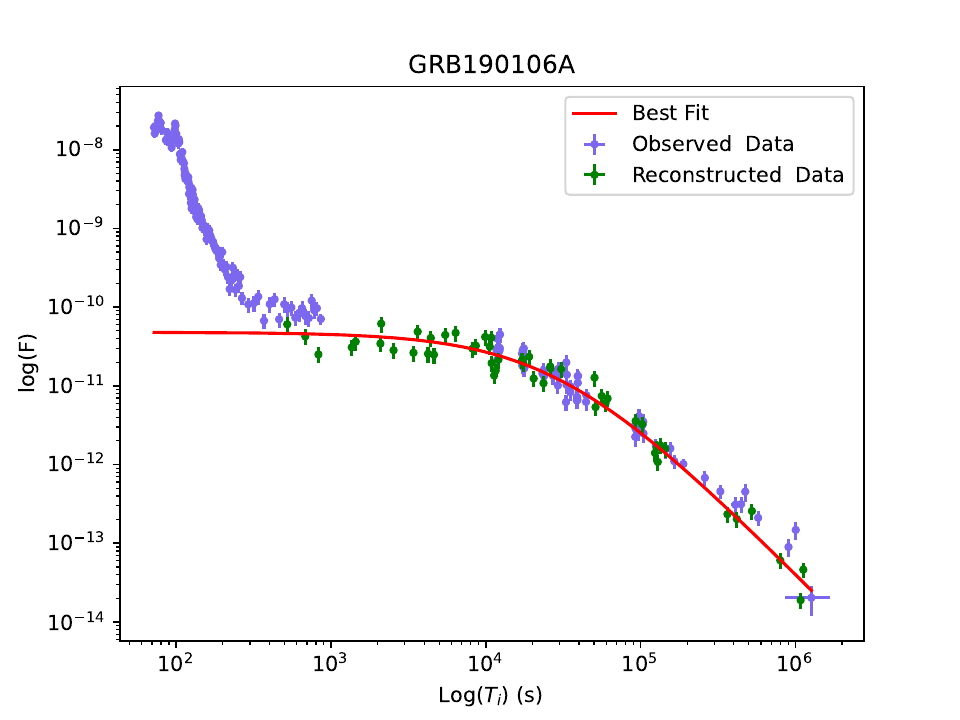}}%
\resizebox{45mm}{!}{\includegraphics[]{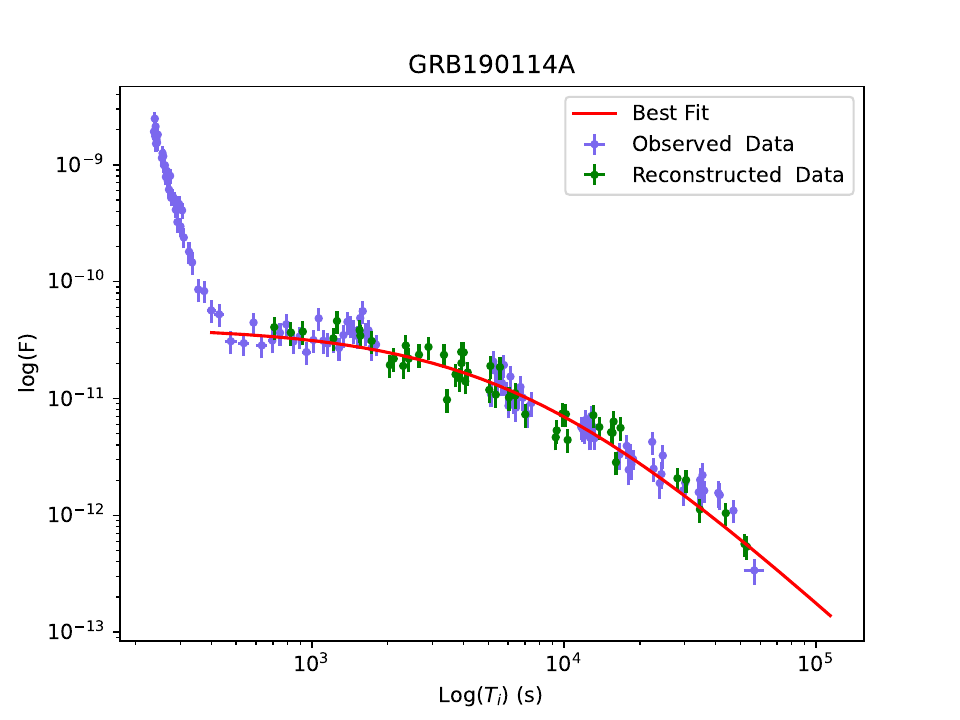}}%
\resizebox{45mm}{!}{\includegraphics[]{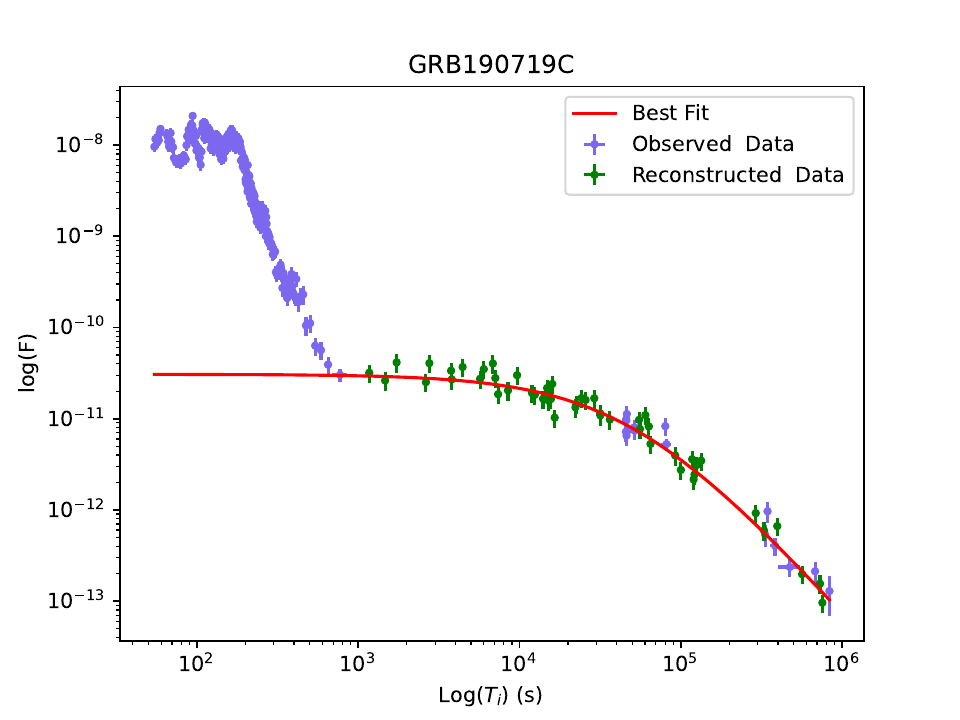}}%
\resizebox{45mm}{!}{\includegraphics[]{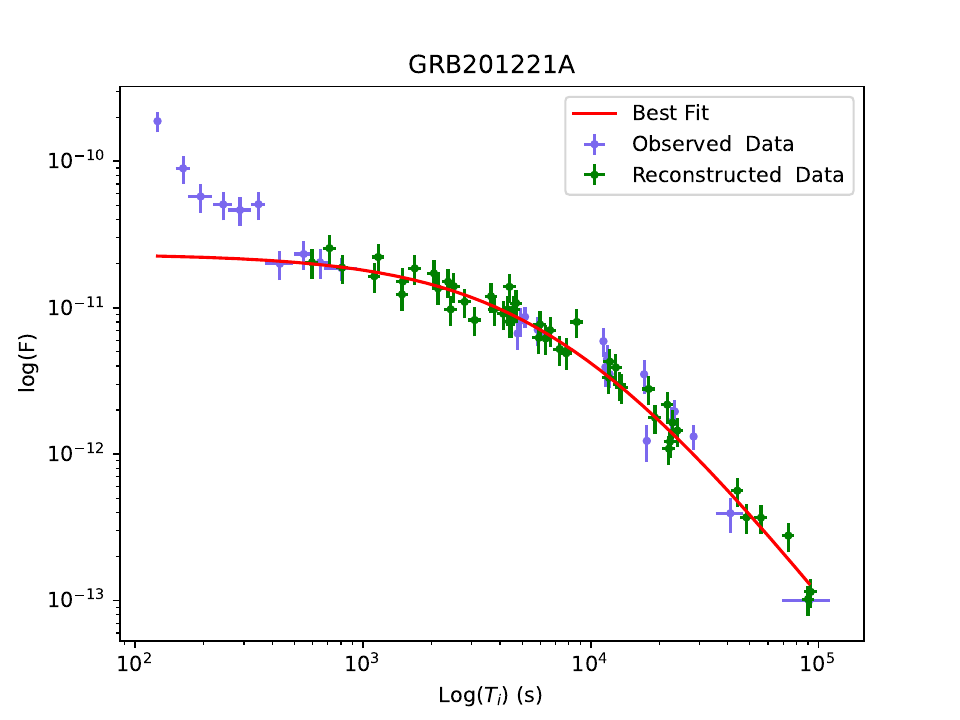}}\\

\resizebox{45mm}{!}{\includegraphics[]{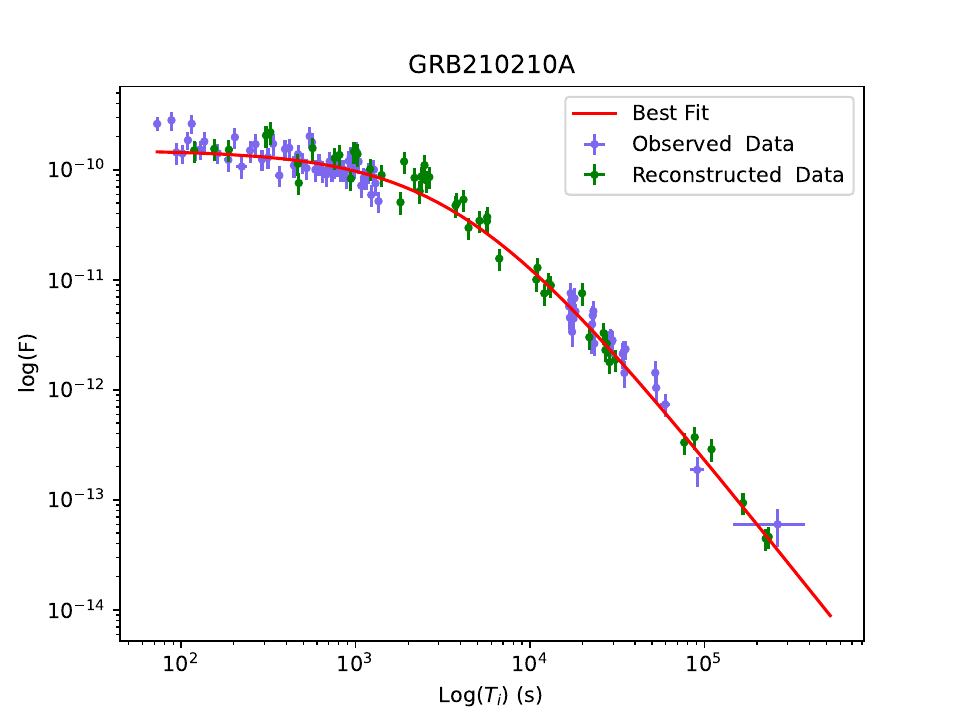}}%
\resizebox{45mm}{!}{\includegraphics[]{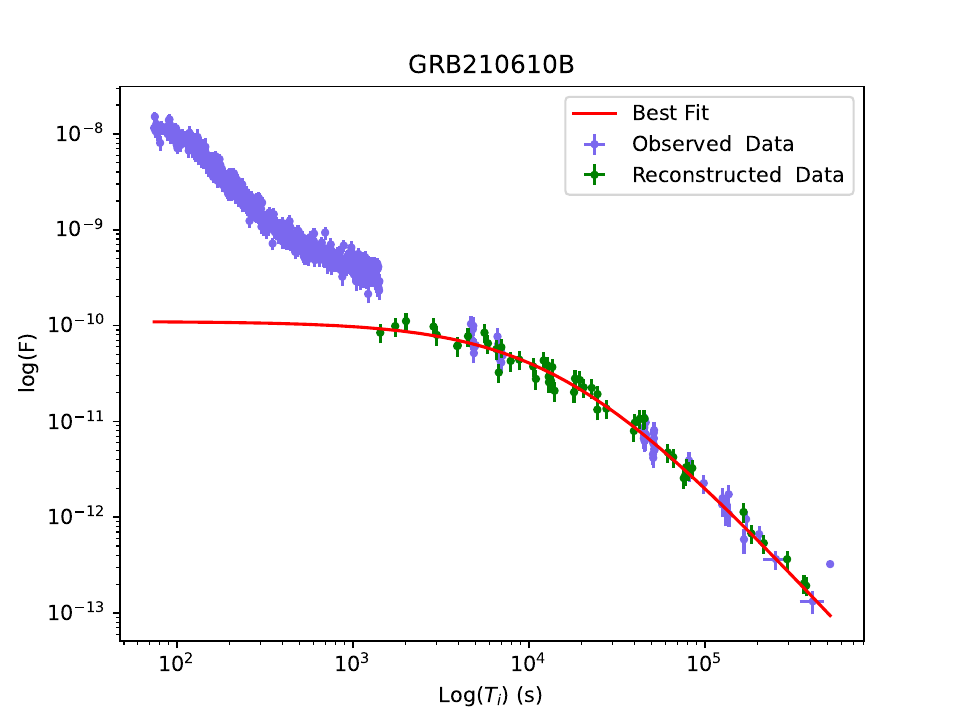}}%
\resizebox{45mm}{!}{\includegraphics[]{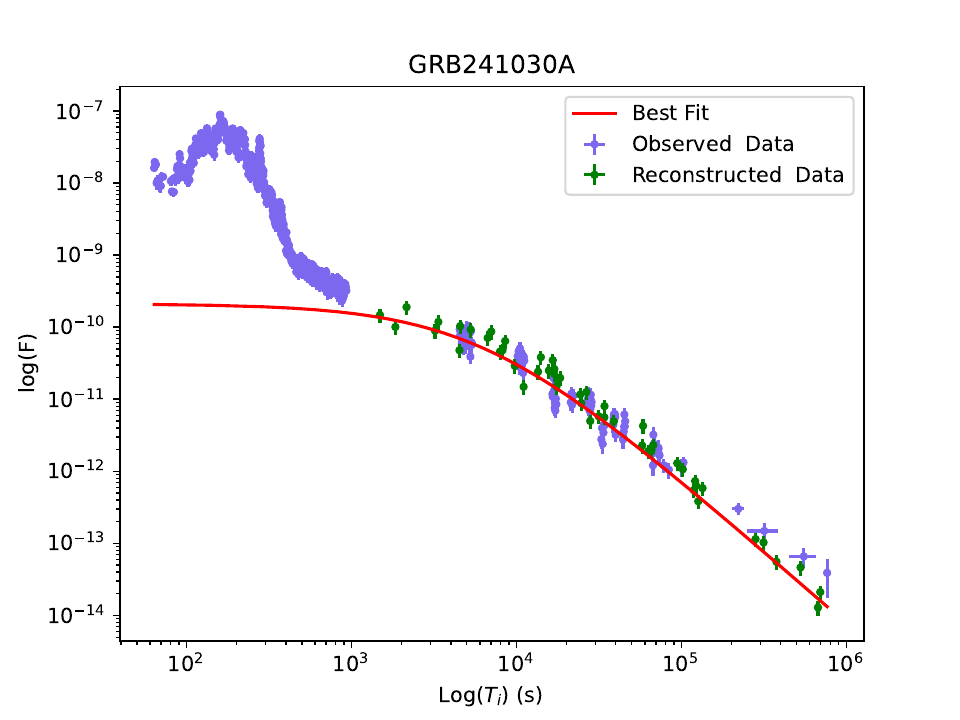}}%
\caption{(Continued)}
\end{figure*}

Perform the aforementioned reconstruction steps 100 times for each GRB, which helps to improve the accuracy and stability of the reconstruction results. By combining the original observational data points, we obtain a new enhanced light curve, referred to as the combined GRB LCs. Subsequently, we fit the combined GRB LCs using Equation (\hyperref[eq:1]{1}). In this way, each GRB has 100 sets of fitting parameters and errors derived from the magnetar model. The uncertainty in the flux of each reconstructed data point is taken as the average of the uncertainties of the flux from the original observational data points of the GRB.

\section{The correlation of GRBs}
\label{section:3}

The luminosity correlations of GRBs offer a valuable means for cosmological observations, enabling us to explore the universe at redshifts reaching approximately $z \sim 8-9$. In the following sections, we will present and calibrate three luminosity correlations of GRBs.

\subsection{Calibrating $L_0-t_b$ correlation}

Dainotti or $L_0-t_b$ correlation is a afterglow emission correlation expressed as

\begin{equation}\label{}
\log_{}{\left ( \frac{L_0}{10^{47} erg/s} \right )}    =a+b\log_{}{ \frac{t_b}{10^3 \left(1+z\right)s}}
\label{eq:5}
\end{equation}
with intercept $a$, slope $b$ and the intrinsic dispersion $\sigma_{int}$.
It relates the X-ray afterglow luminosity $L_0$ to the plateau ending time $t_b$. $L_0$ is related to the plateau flux $F_0$ :
\begin{equation}\label{}
L=\frac{4\pi d_{L}^{2}F}{\left ( 1+z \right )^{1-\beta } }
\label{eq:6}
\end{equation}
where $\beta$ is the  spectral index of the plateau phase and $d_L$ is the luminosity distance. In the flat universe model, the luminosity distance $d_L$ can be written as
 \begin{equation}\label{}
d_L=\frac{c\left ( 1+z \right ) }{H_0} \int_{0}^{z}\frac{dz}{\sqrt{\Omega_m\left ( 1+z \right ) ^3+\Omega_\Lambda} }
\label{eq:7}
\end{equation}
where $\Omega_m$ and $\Omega_\Lambda$ represent the density of matter and dark energy, respectively. Here, we assume that $\Omega_m$ = 0.3, $\Omega_\Lambda$ = 0.7 and $H_0$=70 km/s/Mpc.

The best-fitting results of $a$, $b$ and the intrinsic scatter $\sigma_{int}$ can be obtained by using the log-likelihood function

\begin{equation}\label{}
\begin{aligned}
\mathcal{L}\left ( a,b,\sigma _{int} \right )
&\propto \prod_{i}^{} \frac{1}{\sqrt{\sigma _{int}^2+\sigma_{y_i}^2+{b^2}\sigma_{x_i}^2}}\\
&\times \exp\left [ -\frac{\left ( y_i-a-bx_i \right )^2 }
{2\left ( \sigma _{int}^2+\sigma_{y_i}^2+{b^2}\sigma_{x_i}^2 \right ) }  \right ]
\end{aligned}
\label{eq:8}
\end{equation}

Previously, we fixed the values of $\Omega_m$ and $\Omega_\Lambda$ and calculated the luminosity distance $d_L$ for each GRB. However, when employing correlations to constrain cosmological parameters, a circularity issue emerges. There are now many methods to address this problem, and in this paper, we utilize a method that leverages Gaussian processes (GP) to calibrate these correlations. Specifically, we use the public code GaPP\footnote{https://github.com/carlosandrepaes/GaPP} to perform the GP regression, which offers the advantage of yielding a calibration correlation that is independent of any particular model.

\begin{figure}[ht!]\
\centering
\resizebox{60mm}{!}{\includegraphics[]{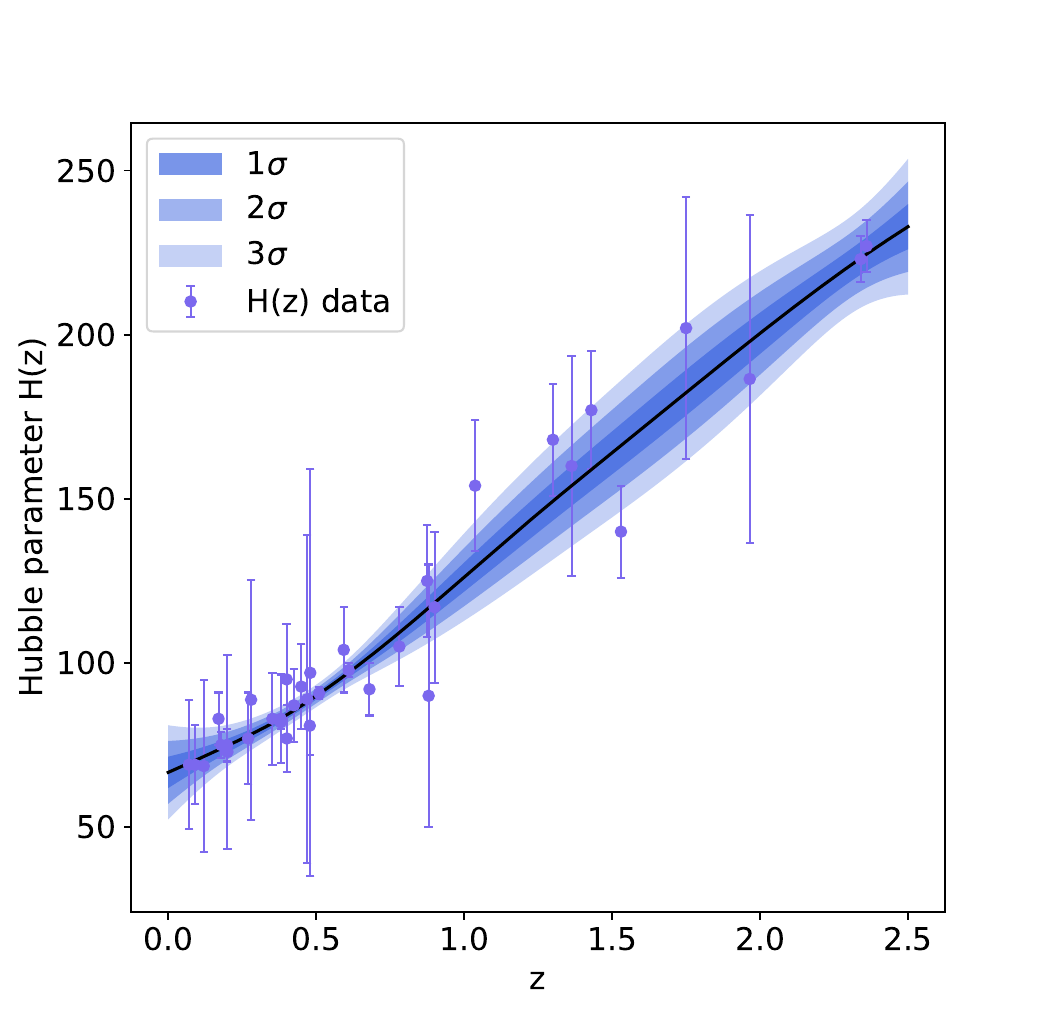}}\\
\caption{The smoothed $H(z)$ function reconstructed using the GP method. The shaded areas represent the 1$\sigma$, 2$\sigma$, and 3$\sigma$ confidence intervals, respectively.}
\label{fig:2}
\end{figure}

First, we calibrate the luminosity distance $d_L$ using the 36 low-redshift $H\left ( z \right )$ data (at $z<2.36$) summarized in \cite{2018ApJ...856....3Y}. We utilize the GP method to calibrate the GRB luminosity distances for those with redshifts less than 2.5 in our data sample. Referring to the methods in \cite{2021MNRAS.507..730H} and \cite{2022ApJ...924...97W}, we perform GP reconstruction for $H(z)$ regarding the luminosity distance of the Hubble parameter $H(z)$,
\begin{equation}\label{}
d_L\left ( z \right ) =c\left ( 1+z \right ) \int_{0}^{z} \frac{dz}{H\left ( z \right ) }
\label{eq:9}
\end{equation}
the integral values of $1/H(z)$ at different redshifts $z$ can be calculated through this continuous function. Figure $\ref{fig:2}$ shows the reconstruction results of the $H(z)$ curve. By combining the above equation, we can obtain the calibrated luminosity distance for each low-redshift GRB. Using Equations (\hyperref[eq:6]{6}), we can obtain the calibrated GRB luminosity. Then, the best-fit parameters for the calibrated correlation were obtained by combining with Equation (\hyperref[eq:5]{5}).

\begin{figure*}[ht!]\
\resizebox{90mm}{!}{\includegraphics[]{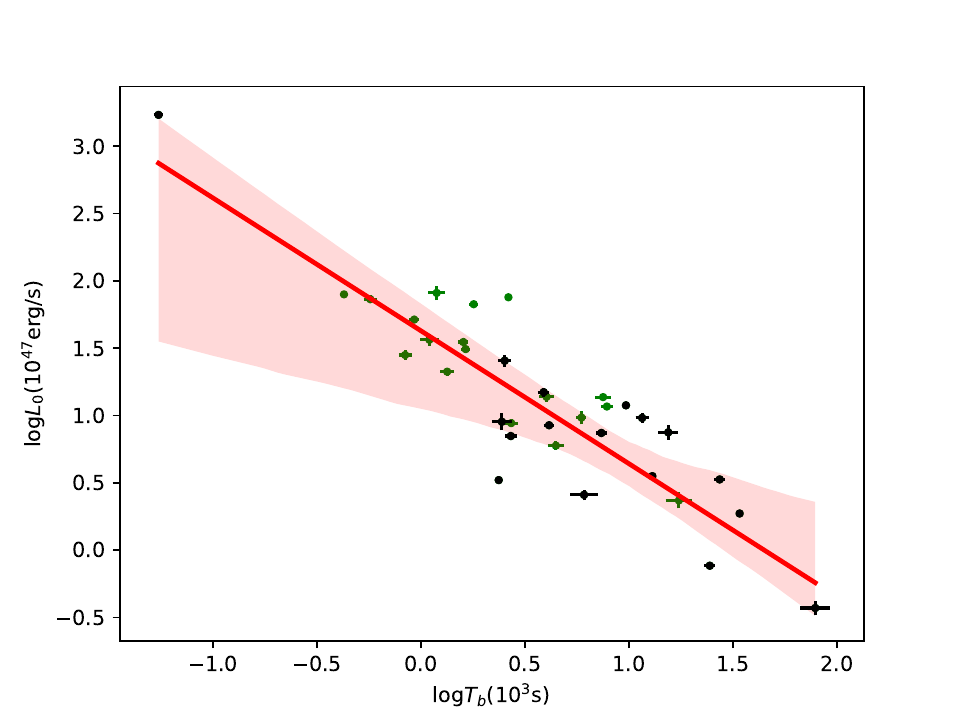}}\resizebox{90mm}{!}{\includegraphics[]{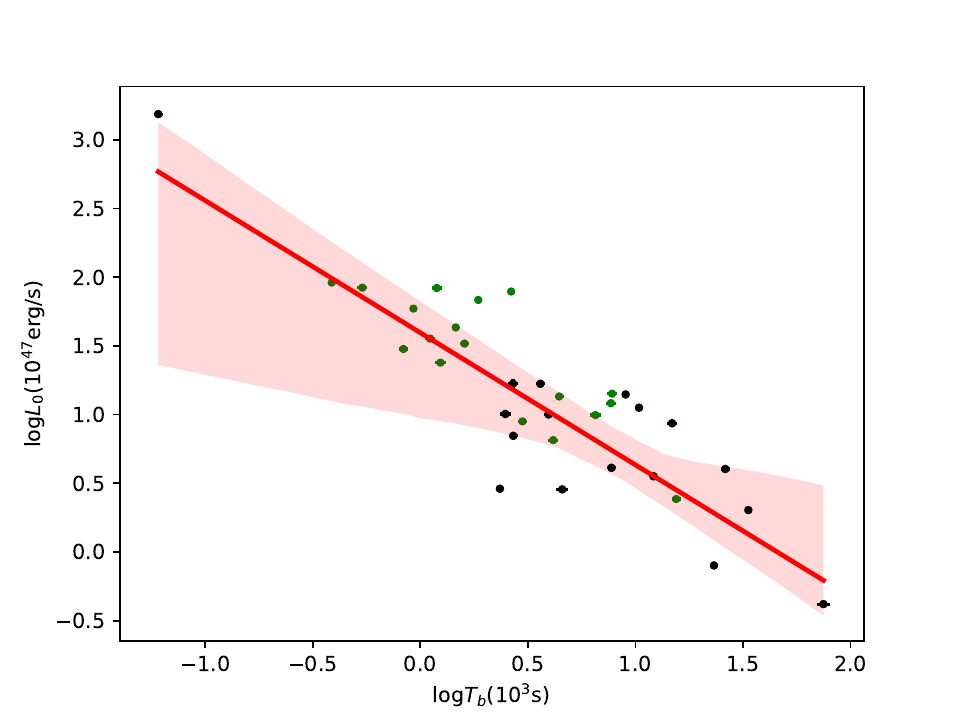}}\\
\caption{The $L_0-t_b$ correlation. The black points are those with $z$ $<$ 2.5 (which can be used to calibrate the luminosity distance $d_L$ using the data from the Figure $\ref{fig:2}$), while the green points represent GRBs with $z$ $>$ 2.5. The left image shows the fitted plot of correlations before reconstruction, while the right image shows the fitted plot of correlations after reconstruction.}
\label{fig:3}
\end{figure*}

Using the public available  emcee package, we obtained the best-fit values for $a,b$ and $\sigma_{int}$. Figure \ref{fig:3} shows the Dainotti correlation. The best-fit results before reconstruction are $a=1.626_{-0.148}^{+0.145}$,  $b=-0.983_{-0.137}^{+0.137}$ and $\sigma_{int}=0.380_{-0.063}^{+0.086}$. The best results after reconstruction are $a=1.594_{-0.155}^{+0.155}$, $b=-0.961_{-0.147}^{+0.153}$ and $\sigma_{int}=0.406_{-0.067}^{+0.093}$. All the errors are within the 1$\sigma$ confidence interval. The detailed fitting results for these parameters are presented in Table \ref{tab:addlabe2}.

Certain selection effects (e.g., the correlated variables $t_b$ and $L_0$ being influenced by the detector's sensitivity threshold) may affect the luminosity correlation. \cite{2013MNRAS.436...82D} investigated the redshift dependence of $t_b$ and $L_0$, finding this correlation to be robust. Furthermore, after removing the redshift dependence of $t_b$ and $L_0$, the intrinsic slope b derived from 101 GRBs was $-1.07_{-0.09}^{+0.14}$ (\citealp{2015MNRAS.451.3898D}), which is consistent within errors with our slope before the reconstruction. Some works also confirmed this correlation (\citealp{2010ApJ...722L.215D, 2011ApJ...730..135D, 2015MNRAS.451.3898D, 2017A&A...600A..98D, 2020ApJ...905L..26D, 2023ApJ...951...63D, 2023MNRAS.518.2201D}; \citealp{2016ApJ...828...36D}; \citealp{2018ApJ...863...50S}; \citealp{2019ApJS..245....1T}; \citealp{2019ApJ...883...97Z}).

Our analysis reveals that, regardless of whether reconstruction is performed, the slope of the relation under consideration remains approximately -1. By incorporating this finding with Equation (\ref{eq:5}), it can be inferred that the product $L_0 \times t_b$ remain constant. This implies that the energy injected by the magnetar per unit time is constant over time. Such a constant energy injection rate suggests that the millisecond magnetar may maintain a stable magnetic field and rotational frequency throughout its evolution. This stability could allow the GRB to continuously release energy, which helps explain some of the observed long durations of GRBs.

\subsection{Calibrating $L_0-t_b-E_{p,i}$ correlation}

$L_0-t_b-E_{p,i}$ correlation is a hybrid prompt\--{}afterglow emission correlation expressed as

\begin{equation}\label{}
\log_{}{\left ( \frac{L_0}{10^{47}\space erg/s} \right ) }   =a'+b'\log_{}{ \frac{t_b}{10^3 \left(1+z\right)s}}+c'\log_{}{ \frac{E_{p,i}}{keV}}
\label{eq:10}
\end{equation}
Here, $E_{p,i} = E_{p,obs} \times (1 + z)$ is the spectral peak energy with the observed peak energy $E_{p,obs}$.

\begin{figure*}[ht!]\
\resizebox{90mm}{!}{\includegraphics[]{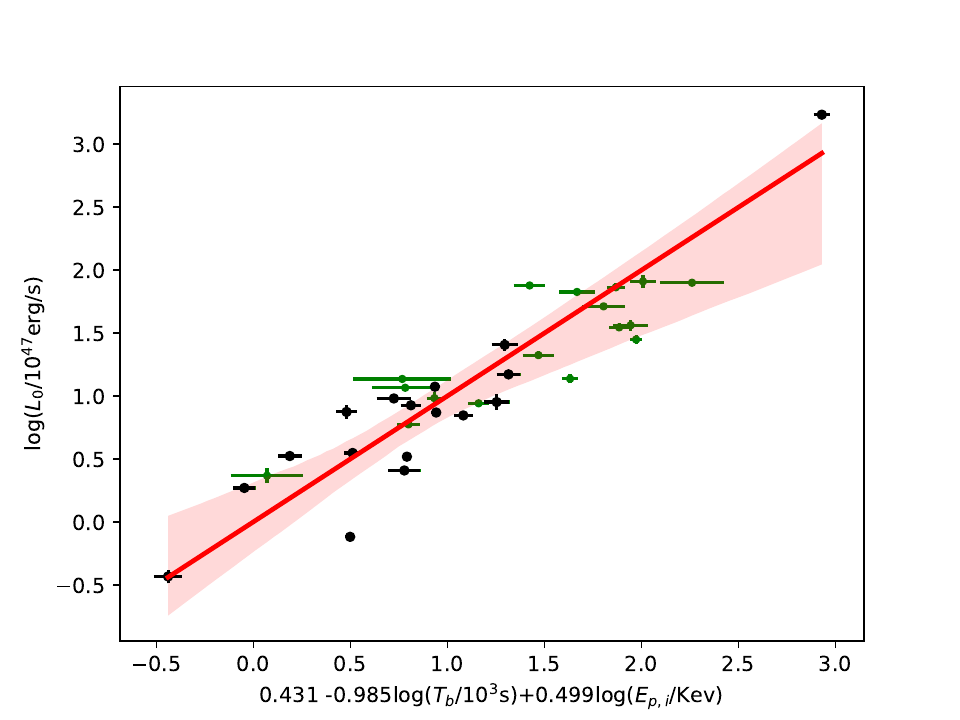}}\resizebox{90mm}{!}{\includegraphics[]{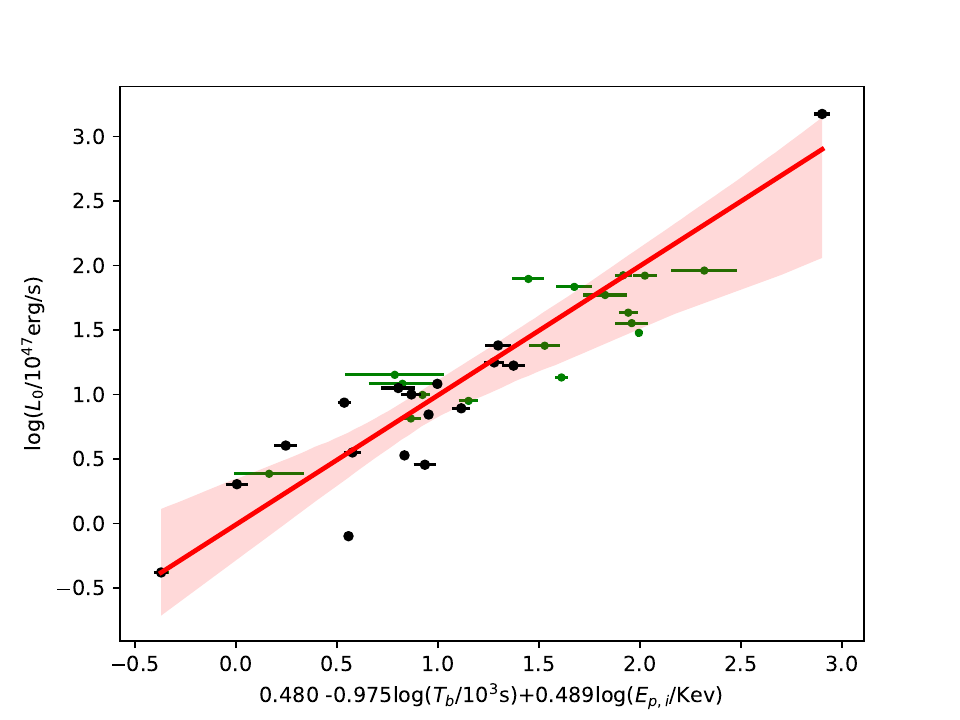}}\\
\caption{The $L_0-t_b-E_{p,i}$ correlation with $T_b=t_b/(1+z)$. The black points are those with $z$ $<$ 2.5 ( which can be used to calibrate the luminosity distance $d_L$), while the green points represent GRBs with $z$ $>$ 2.5. The left image shows the fitted plot of correlations before reconstruction, while the right image shows the fitted plot of correlations after reconstruction.}
\label{fig:4}
\end{figure*}

\begin{figure*}[ht!]\
\resizebox{90mm}{!}{\includegraphics[]{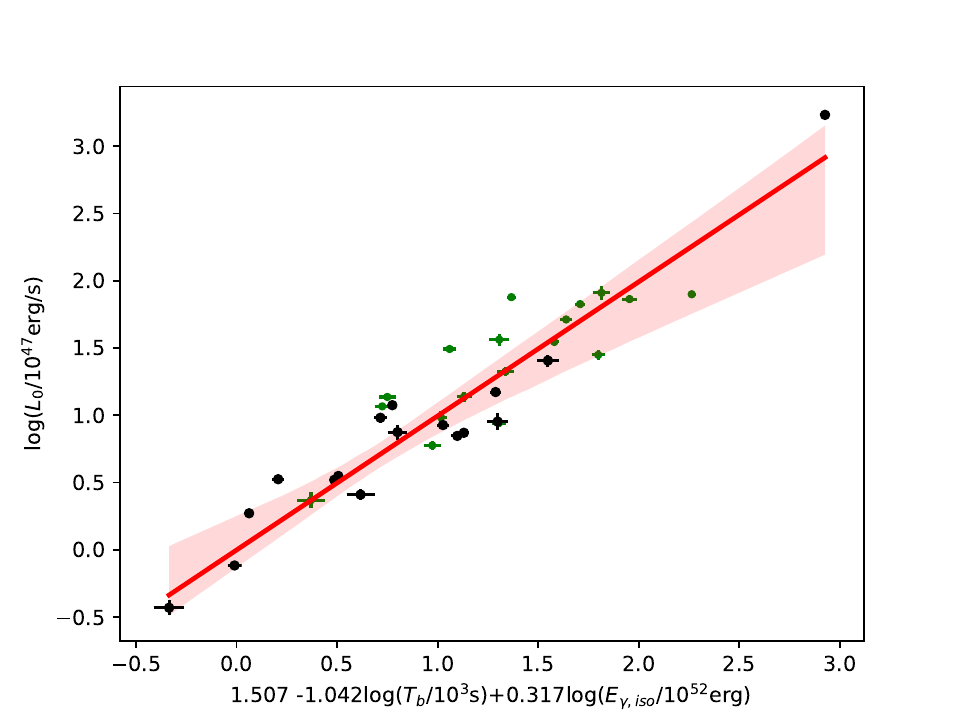}}\resizebox{90mm}{!}{\includegraphics[]{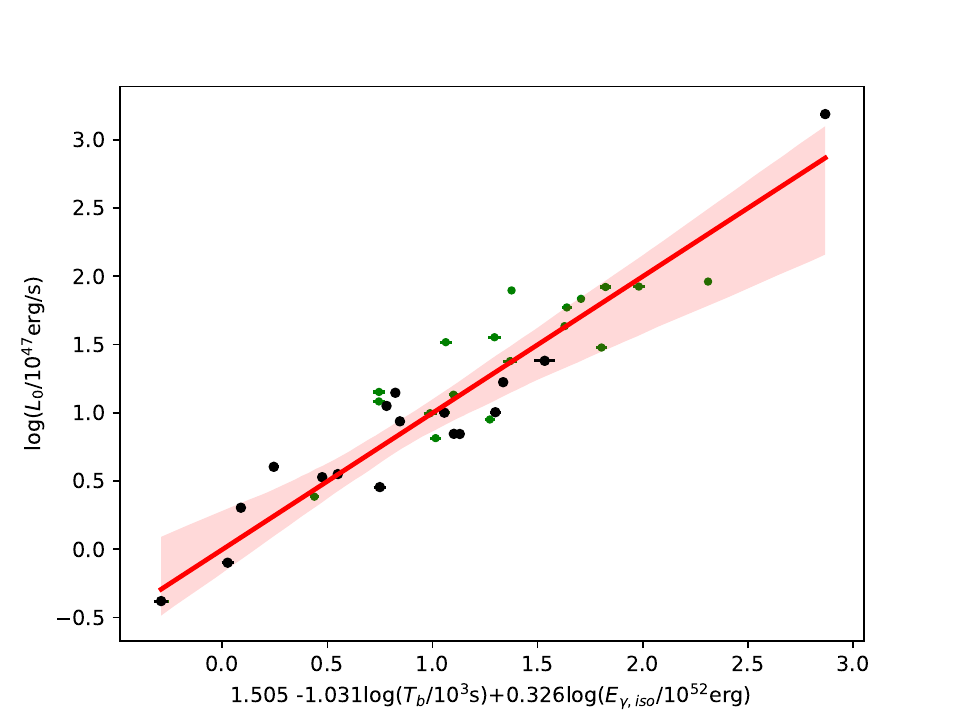}}\\
\caption{The $L_0-t_b-E_{\gamma,\mathrm{iso}}$ correlation with $T_b=t_b/(1+z)$. The black points are those with $z$ $<$ 2.5 (which can be used to calibrate the luminosity distance $d_L$), while the green points represent GRBs with $z$ $>$ 2.5. The left image shows the fitted plot of correlations before reconstruction, while the right image shows the fitted plot of correlations after reconstruction.}
\label{fig:5}
\end{figure*}

We collected data from the following literature: \cite{2018ApJ...863...50S}, \cite{2020MNRAS.492.1919M}, \cite{2021ApJ...920..135X} and \cite{2021MNRAS.508...52L}. For GRB 210610B and GRB 241030A, we referred to the results provided by KONUS-WIND on GCN (\texttt{https://gcn.nasa.gov/circulars}) for its $E_{p,i}$. Unfortunately, for GRB 190114A, we were unable to find the value of its peak energy. Consequently, this particular burst was excluded from our analysis.

The best-fitting results of $a$, $b$, $c$, and the intrinsic scatter $\sigma_{int}$ can be obtained by using the log-likelihood function

\begin{equation}\label{}
\begin{aligned}
\mathcal{L}\left ( a,b,c,\sigma _{int} \right )
&\propto \prod_{i}^{}  \frac{1}{\sqrt{\sigma _{int}^2+\sigma_{y_i}^2+{b^2}\sigma_{x_{1,i}}^2+{c^2}\sigma_{x_{2,i}}^2}}\\
&\times \exp\left [ -\frac{\left ( y_i-a-bx_{1,i}-cx_{2,i} \right )^2 }
{2\left ( \sigma _{int}^2+\sigma_{y_i}^2+{b^2}\sigma_{x_{1,i}}^2+c^2\sigma_{x_{2,i}}^2 \right ) }  \right ]
\end{aligned}
\label{eq:11}
\end{equation}

Figure \ref{fig:4} shows the $L_0-t_b-E_{p,i}$ correlation. Prior to reconstruction, the best-fit values for the parameters are determined to be $a=0.431_{-0.515}^{+0.528}$, $b=-0.985_{-0.116}^{+0.118}$, $c=0.499_{-0.210}^{+0.203}$, and $\sigma_{int}=0.321_{-0.058}^{+0.079}$. Following the reconstruction process, the optimal parameter estimates are updated to $a=0.480_{-0.525}^{+0.516}$, $b=-0.975_{-0.125}^{+0.124}$, $c=0.489_{-0.211}^{+0.213}$, and $\sigma_{int}=0.332_{-0.057}^{+0.081}$. All the errors are within the 1$\sigma$ confidence interval. A comprehensive summary of the fitting results for these parameters can be found in Table \ref{tab:addlabe2}.

\subsection{Calibrating $L_0-t_b-E_{\gamma,\mathrm{iso}}$ correlation}

$L_0-t_b-E_{\gamma,\mathrm{iso}}$ correlation is a hybrid prompt--afterglow emission correlation expressed as
\begin{equation}\label{}
\log_{}{\left ( \frac{L_0}{10^{47}\space erg/s} \right ) }   =a''+b''\log_{}{ \frac{t_0}{10^3 \space s}}+c''\log_{}{ \frac{E_{\gamma,\mathrm{iso}}}{10^{52}\space erg}}
\label{eq:12}
\end{equation}

The isotropic energy of the prompt emission is
\begin{equation}\label{}
E_{\gamma,\mathrm{iso}}'=\frac{4\pi d_L^2S}{1+z}
\label{eq:13}
\end{equation}
where $S$ is the observed bolometric fluence. More detailed discussions about the K-correction process can be found in, e.g., \cite{1993ApJ...413..281B}, \cite{2013EAS....61..381S}, \cite{2023ApJ...958...74T} and \cite{2023ApJ...953...58L}.

We collected data from the following literature: \cite{2012A&A...538A.134X}, \cite{2023ApJ...943..126D} and \cite{2018ApJ...863...50S}. For GRB 210610B, GRB 220117A and GRB 241030A, we referred to the results provided by KONUS-WIND on GCN for its $E_{\gamma,\mathrm{iso}}$. The best fit results for $a$, $b$, and $c$, as well as the intrinsic scatter $\sigma_{int}$, can be obtained using the logarithmic likelihood function, which has a form similar to Equation (\hyperref[eq:12]{12}).

Figure \ref{fig:5} shows the $L_0-t_b-E_{\gamma,\mathrm{iso}}$ correlation. The best-fit results before reconstruction are constrained to $a=1.507_{-0.099}^{+0.100}$, $b=-1.042_{-0.090}^{+0.093}$, $c=0.317_{-0.072}^{+0.073}$ and $\sigma_{int}=0.245_{-0.045}^{+0.059}$. After reconstruction, the parameters are determined to $a=1.505_{-0.101}^{+0.102}$, $b=-1.031_{-0.099}^{+0.095}$, $c=0.326_{-0.077}^{+0.077}$ and $\sigma_{int}=0.259_{-0.043}^{+0.063}$. All the errors are within the 1$\sigma$ confidence interval. The fitting results of the parameters are shown in Table \ref{tab:addlabe2}.

\section{Constraints on the cosmological parameters}
\label{section:4}
In the flat $\Lambda$CDM model, the distance modulus can be written as
\begin{equation}\label{}
\mu_{th}=5\log_{}{\frac{d_L}{\text{Mpc}}} +25=5\log_{}{\frac{d_L}{\text{cm}}}-97.45
\label{eq:14}
\end{equation}

By utilizing Equation (\hyperref[eq:6]{6}), one can express $d_L$ in terms of $L_0$, and the observed distance modulus can be reformulated as:
\begin{equation}\label{}
\begin{aligned}
\mu_{obs}=\frac{5}{2}
\left[ \log L_0-\log_{}{\frac{4\pi F}{(1+z)^{1-\beta}}}\right]-97.45
\end{aligned}
\label{eq:15}
\end{equation}
The $L_0$ for the two-parameter and three-parameter correlations are determined by Equation (\hyperref[eq:5]{5}) and Equation (\hyperref[eq:12]{12}), respectively.

The uncertainty of $\mu_{obs}$ can be described as:
\begin{equation}\label{}
\begin{aligned}
\sigma_{obs}=\frac{5}{2}
\left [ {\sigma_{int}^2+\sigma_a^2+\sigma_b^2 \left ( \log_{}{ \frac{t_b}{1+z} }-3 \right )^2 }
\right.\\
\left.
 {+b^2\left ( \frac{\sigma_{t_b}}{t_b\ln_{}{10} }  \right )^2+\left ( \frac{\sigma_{F_0}}{F_0\ln_{}{10} } \right )^2 }\right ]^{1/2}\\
\end{aligned}
\label{eq:16}
\end{equation}
and
\begin{equation}\label{}
\begin{split}
\sigma_{obs}&=\frac{5}{2}[ \sigma_{int}^2+\sigma_a^2+\sigma_b^2 \left ( \log_{}{ \frac{t_b}{1+z} }-3 \right )^2 \\
&+ b^2\left ( \frac{\sigma_{t_b}}{t_b\ln_{}{10} }  \right )^2 +c^2\left ( \frac{\sigma_{E_{p,i}}}{E_{p,i}\ln_{}{10} } \right )^2  \\
&+\left ( \frac{\sigma_{F_0}}{F_0\ln_{}{10} } \right )^2+\sigma_c^2\log_{}{E_{p,i}^2}  ]^{1/2}
\end{split}
\label{eq:17}
\end{equation}
Here, we take the $E_{p,i}$ as the example. For the $L_0-t_b-E_{\gamma,\mathrm{iso}}$ correlation, one only needs to replace $E_{p,i}$ with $E_{\gamma,\mathrm{iso}}$. The calculation of each error in $\sigma_{i}$ is performed using $\sqrt{\frac{(\sigma_u^2+\sigma_d^2)^2}{2} } $, where $\sigma_u$ ($\sigma_d$) represents the upper (lower) error of that value.

To calculate the theoretical distance modulus $\mu_{th}$, we first need to establish the theoretical luminosity distance.
\begin{equation}\label{}
d_L =
\begin{cases}
\frac{c( 1+z  ) }{H_0}( -\Omega_k  )^{-\frac{1}{2} }
\sin \left [  ( -\Omega_k   )^{-\frac{1}{2} } \int_{0}^{z}\frac{dz}{E(z)}   \right ]
 &,{\Omega_k< 0}\\
\frac{c( 1+z  ) }{H_0} \int_{0}^{z}\frac{dz}{E(z)}
&,{\Omega_k=  0}\\
\frac{c( 1+z  ) }{H_0}  \Omega_k  ^{-\frac{1}{2} }
\sinh \left [ \Omega_k  ^{-\frac{1}{2} } \int_{0}^{z}\frac{dz}{E(z)}   \right ]
&,{\Omega_k>  0}
\end{cases}
\label{eq:18}
\end{equation}
where $\Omega_k$ denotes spatial curvature. $E(z)$ can be written as
\begin{equation}\label{}
E(z)=\sqrt{\Omega_m(1+z)^3+(1-\Omega_m-\Omega_\Lambda)(1+z)^2+\Omega_\Lambda}
\label{eq:19}
\end{equation}
The best-fitting parameters can be obtained by minimizing $\chi^2$
\begin{equation}\label{}
\chi^2=\sum_{j=1}^{35}  \frac{ \left [ \mu_{obs}\left ( z \right ) -\mu_{th}\left ( \Omega_i,z \right ) \right ]^2 }{\sigma_{obs}^2}
\label{eq:20}
\end{equation}
where $\Omega_i$ represents the cosmological parameters that needs to be constrained: $\Omega_m$ for a flat universe, and both $\Omega_m$ and $\Omega_\Lambda$ for a non-flat universe.

For constraining cosmological parameters, we incorporated data from SNe Ia. Prior to performing any reconstruction, our analysis of the flat $\Lambda$CDM model using the $L_0-t_b$ correlation yielded a best-fit value for the matter density parameter of $\Omega_m = 0.290_{-0.007}^{+0.008}$. When considering a non-flat model, the constraints obtained are $\Omega_m = 0.331_{-0.033}^{+0.033}$ and $\Omega_\Lambda = 0.762_{-0.041}^{+0.040}$. Using calibrated $L_0-t_b-E_{p,i}$ correlation and calibrated $L_0-t_b-E_{\gamma,\mathrm{iso}}$ correlation to constrain the flat $\Lambda$CDM model results in $\Omega_m$ being consistent-both are $0.290_{-0.008}^{+0.008}$. The constraints of $\Omega_m$ and $\Omega_\Lambda$ values under the non-flat $\Lambda$CDM model are different for the two, which are $\Omega_m = 0.335_{-0.035}^{+0.036}$ and $\Omega_\Lambda = 0.767_{-0.044}^{+0.044}$ ($L_0-t_b-E_{p,i}$ correlation) and $\Omega_m = 0.328_{-0.030}^{+0.030}$ and $\Omega_\Lambda = 0.758_{-0.037}^{+0.037}$ ($L_0-t_b-E_{\gamma,\mathrm{iso}}$ correlation). After randomly reconstructing each GRB samples (with 50 reconstruction points), the cosmological parameters obtained by constraining the flat $\Lambda$CDM model using three different correlations are consistent, all yielding a result of $\Omega_m = 0.290_{-0.008}^{+0.008}$. Compared to the value of $\Omega_m$ before reconstruction, it has hardly changed. The reconstructed constraints on the cosmological parameters for the non-flat $\Lambda$CDM model yield the following results: for the $L_0-t_b$ correlation, $\Omega_m = 0.333_{-0.033}^{+0.033}$ and $\Omega_\Lambda = 0.763_{-0.042}^{+0.041}$, for the $L_0-t_b-E_{p,i}$ correlation, $\Omega_m = 0.334_{-0.035}^{+0.036}$ and $\Omega_\Lambda = 0.766_{-0.043}^{+0.044}$, and for the $L_0-t_b-E_{\gamma,\mathrm{iso}}$ correlation, $\Omega_m = 0.329_{-0.030}^{+0.032}$ and $\Omega_\Lambda = 0.759_{-0.039}^{+0.038}$.

\begin{figure*}[ht!]\
\resizebox{80mm}{!}{\includegraphics[]{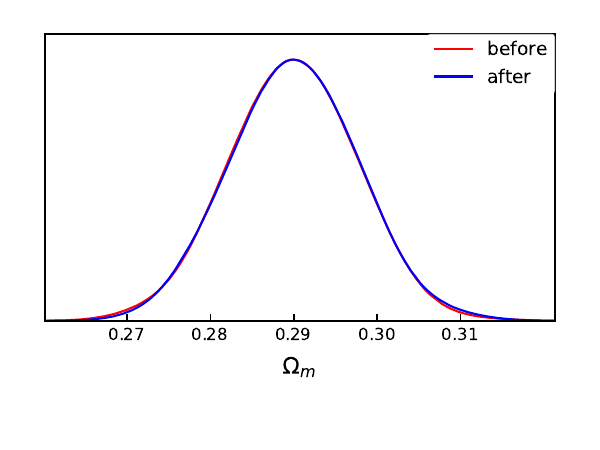}}\resizebox{80mm}{!}{\includegraphics[]{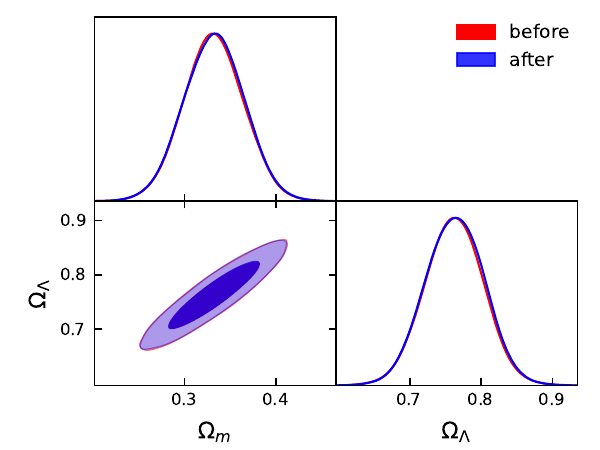}}\\
\caption{The results of cosmological parameters using the $L_0-t_b$ correlation. The left (right) image represents the flat (non-flat) $\Lambda$CDM model. The red (blue) line indicates the fitting results of cosmological parameters before (after) reconstruction.}
\label{fig:6}
\end{figure*}

\begin{figure*}[ht!]\
\resizebox{80mm}{!}{\includegraphics[]{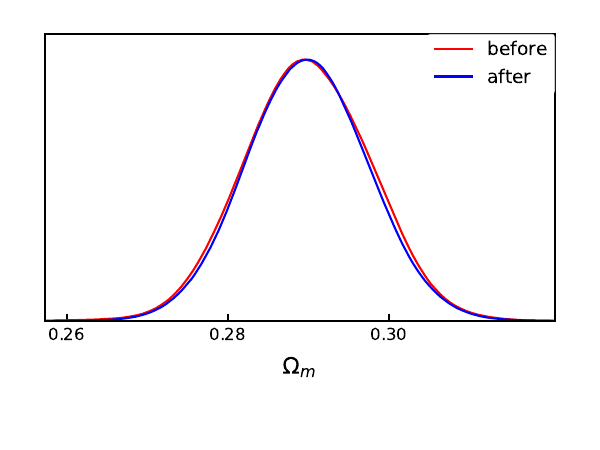}}\resizebox{80mm}{!}{\includegraphics[]{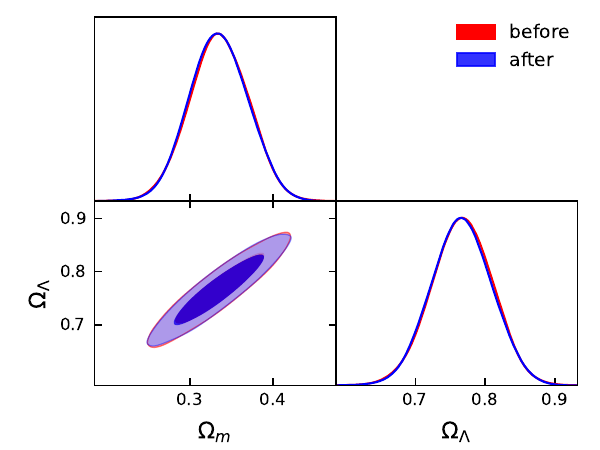}}\\
\caption{The results of cosmological parameters using the $L_0-t_b-E_{p,i}$ correlation. The left (right) image represents the flat (non-flat) $\Lambda$CDM model. The red (blue) line indicates the fitting results of cosmological parameters before reconstruction.}
\label{fig:7}
\end{figure*}

\begin{figure*}[ht!]\
\resizebox{80mm}{!}{\includegraphics[]{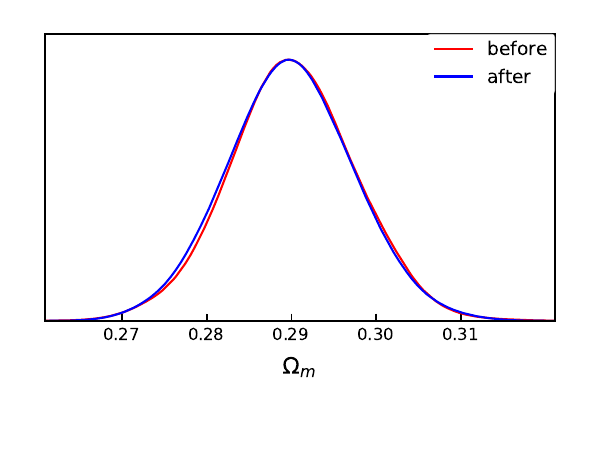}}\resizebox{80mm}{!}{\includegraphics[]{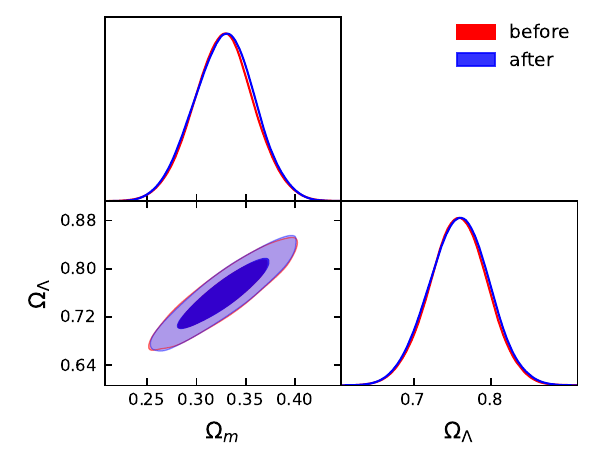}}\\
\caption{The results of cosmological parameters using the $L_0-t_b-E_{\gamma,\mathrm{iso}}$ correlation. The left (right) image represents the flat (non-flat) $\Lambda$CDM model. The red (blue) line indicates the fitting results of cosmological parameters before (after) reconstruction.}
\label{fig:8}
\end{figure*}

Figures \ref{fig:6}, \ref{fig:7}, and \ref{fig:8}, along with Table \ref{tab:addlabe3}, present the results and corresponding images that depict the reconstruction of the constrained cosmological parameters, both prior to and following the reconstruction process.

\section{Results}
\label{section:5}

The primary objective of this study is to utilize a reconstruction method to fill in the gaps present in GRBs LCs. By doing so, we aim to explore how this approach impacts the constraints placed on cosmological parameters. To facilitate this analysis, we have defined the standard to evaluate the variation of uncertainty. The error fraction is defined as
\begin{equation}\label{}
EF_x=\left | \frac{\Delta x}{x}  \right |
\label{eq:21}
\end{equation}
The variable $x$ represents  the best-fit value of a fitting parameter, $\Delta x$ indicates the fitting value of the parameter, and $EF_x$ denotes the error score of this fitting parameter.

Table \ref{tab:addlabe4} shows the error fractions of cosmological parameters constrained by three different correlations. $EF'$ represents the error fractions of the fitting parameter after reconstruction.

We use the percentage reduction of the error fraction to compare and analyze the improvement of the reconstruction method in constraining cosmological parameters, defined by the following formula
\begin{equation}\label{}
\%_{REDU}=\frac{|EF_{before}|-|EF_{after}|}{|EF_{before}|}
\label{eq:22}
\end{equation}

The results of $\%_{REDU}$ are also presented in Table \ref{tab:addlabe4}. A more negative $\%_{REDU}$ indicates that the uncertainty of the parameter after reconstruction has been reduced more. We find that for the flat $\Lambda$CDM model, the $\%_{\Omega_m}$ of the $L_0-t_b$ correlation is positive, while the other two correlations show no change before and after reconstruction. In simpler terms, reconstructing 50 points of the X-ray LCs for our GRB samples did not lead to significant alterations in the constraints on the cosmological parameters. For the $L_0-t_b$ correlation, the precision of the cosmological parameters was actually worse after reconstruction compared to before. For the non-flat $\Lambda$CDM model, only the $L_0-t_b-E_{p,i}$ correlation showed an improvement in the precision of the cosmological parameter constraints after reconstruction. For the other two correlations, reconstruction did not enhance the precision of the cosmological parameters for the non-flat model.

Although the improvement in the precision of cosmological parameters after reconstruction is relatively minor (with a 6.25\% improvement in the precision of $\Omega_m$ in the flat $\Lambda$CDM model and a 1.01\% improvement in the constraint on $\Omega_\Lambda$ in the non-flat $\Lambda$CDM model through $L_0$-$t_b$-$E_{p,i}$ correlation analysis), our work is characterized by the following features:

\begin{itemize}

\item Compared to the sample data from \cite{2022ApJ...924...97W}, we add 4 GRBs that likely exhibit X-ray afterglow light curve plateaus due to magnetar energy injection.

\item We use a stochastic reconstruction method based on the magnetar model functional form to reconstruct the LCs of GRBs. Unlike \cite{2023ApJS..267...42D}, we do not use empirical functions (like the BPL model or the W07 model); instead, we employ the magnetar model related to the physical mechanisms of GRBs for the reconstruction of LCs, and we use the reconstructed LCs to constrain cosmological models.

\item In the fitting of the $L_0-t_b$ correlation, we found that the slope is close to 1. This finding implies that the energy injection from the magnetar remains relatively constant throughout the process.

\item Compared to previous similar cosmological studies, we reconstruct the afterglow LCs of GRBs, filling in the temporal gaps in the GRB LCs and improving the accuracy of the plateau parameters.
\end{itemize}

\section{Discussion and Conclusion}
\label{section:6}

We employ a stochastic reconstruction method to reconstruct the LCs of 35 GRBs with plateau features. Using the reconstructed plateau parameters, we fit the $L_0-t_b$ correlation, the $L_0-t_b-E_{p,i}$ correlation, and the $L_0-t_b-E_{\gamma,\mathrm{iso}}$ correlation.

\begin{itemize}
\item For the $L_0-t_b$ correlation, the best-fit result before reconstruction are $a=1.626_{-0.148}^{+0.145}$,  $b=-0.983_{-0.137}^{+0.137}$ and $\sigma_{int}=0.380_{-0.063}^{+0.086}$; the best-fit result after reconstruction are $a=1.594_{-0.155}^{+0.155}$, $b=-0.961_{-0.147}^{+0.153}$ and $\sigma_{int}=0.406_{-0.067}^{+0.093}$.

\item For the $L_0-t_b-E_{p,i}$ correlation, the best-fit result before reconstruction are $a=0.431_{-0.515}^{+0.528}$, $b=-0.985_{-0.116}^{+0.118}$, $c=0.499_{-0.210}^{+0.203}$ and $\sigma_{int}=0.321_{-0.058}^{+0.079}$; the best results after reconstruction are $a=0.480_{-0.525}^{+0.516}$, $b=-0.975_{-0.125}^{+0.124}$, $c=0.489_{-0.211}^{+0.213}$ and $\sigma_{int}=0.332_{-0.057}^{+0.081}$.

\item For the $L_0-t_b-E_{\gamma,\mathrm{iso}}$ correlation, the best-fit result before reconstruction are $a=1.507_{-0.099}^{+0.100}$, $b=-1.042_{-0.090}^{+0.093}$, $c=0.317_{-0.072}^{+0.073}$ and $\sigma_{int}=0.245_{-0.045}^{+0.059}$; the best results after reconstruction are $a=1.505_{-0.101}^{+0.102}$, $b=-1.031_{-0.099}^{+0.095}$, $c=0.326_{-0.077}^{+0.077}$ and $\sigma_{int}=0.259_{-0.043}^{+0.063}$.
\end{itemize}

One noteworthy phenomenon is the observed difference in the intrinsic scatter ($\sigma_{int}$) before and after light curve reconstruction. Taking the $L_0-t_b$ correlation as an example, the pre-reconstruction $\sigma_{int}$ value is $0.380_{-0.063}^{+0.086}$ (error fraction: 0.198), while the post-reconstruction value becomes $0.406_{-0.067}^{+0.093}$ (error fraction: 0.200). They are consistent within the 1$\sigma$ confidence interval, indicating that although minor fluctuations exist in the intrinsic dispersion.

We attribute this phenomenon to several potential factors. Firstly, the limited sample size - our study contains only 35 GRBs, resulting in sparse data points for luminosity correlation fitting that directly impacts $\sigma_{int}$ measurement precision. We anticipate that next-generation space observatories like SVOM (Space Multi-band Variable Object Monitor; \citealp{2016arXiv161006892W}), Einstein Probe (\citealp{2015arXiv150607735Y}), and THESEUS (Transient High-Energy Sky and Early Universe Surveyor; \citealp{2018AdSpR..62..191A}), combined with ground-space multi-messenger facilities, will substantially increase high-quality GRB samples, thereby effectively reducing $\sigma_{int}$ values and uncertainties. Secondly, GRB parameters may be influenced by multiple physical mechanisms. Recent work by \cite{2024arXiv241220091M} classifies GRBs into 4 distinct types based on light curve characteristics, each requiring different reconstruction approaches. However, we uniformly applied the magnetar model to all GRB LCs. The stochastic nature of reconstruction sampling may introduce systematic offsets in plateau parameters by neglecting GRB diversity features, consequently affecting the original $L_0-t_b$ correlation. Finally, we consistently adopted 50 reconstructed data points per GRB LC. This methodological selection effect may introduce additional scatter.

By utilizing a reconstructed sample of 35 GRBs alongside data from Type Ia supernovae, we conducted a joint analysis to constrain both the flat and non-flat $\Lambda$CDM models. This analysis allowed us to determine the best-fit values for the cosmological parameters following
the reconstruction process.
Specifically, for the flat $\Lambda$CDM model, we found $\Omega_m=0.290_{-0.008}^{+0.008}$ using the $L_0-t_b$ correlation, $\Omega_m=0.290_{-0.008}^{+0.007}$ using the $L_0-t_b-E_{p,i}$ correlation, and $\Omega_m=0.290_{-0.008}^{+0.007}$ using the $L_0-t_b-E_{\gamma,\mathrm{iso}}$ correlation. For the non-flat $\Lambda$CDM model, the results are $\Omega_m=0.333_{-0.033}^{+0.033}$ and $\Omega_\Lambda=0.763_{-0.042}^{+0.041}$ in the $L_0-t_b$ correlation, $\Omega_m=0.334_{-0.035}^{+0.036}$ and $\Omega_\Lambda=0.766_{-0.043}^{+0.044}$ in the $L_0-t_b-E_{p,i}$ correlation, and $\Omega_m=0.329_{-0.030}^{+0.032}$ and $\Omega_\Lambda=0.759_{-0.039}^{+0.038}$ in the $L_0-t_b-E_{\gamma,\mathrm{iso}}$ correlation.

For flat $\Lambda$CDM model, the precision of $\Omega_m$ improves by 6.25\% in the $L_0-t_b-E_{p,i}$ correlation. For the non-flat $\Lambda$CDM model, the precision of $\Omega_m$ improves by 0.60\% in the $L_0-t_b$ correlation, while the precision of $\Omega_\Lambda$ improves by 1.01\% in the $L_0-t_b-E_{p,i}$ correlation. We speculate that the number of data points observed during the afterglow phase and the amount of gaps are not the key factors affecting the precision of cosmological parameters. Instead, we anticipate that as the sample size of GRBs at high redshifts increases in the future, there will be a meaningful improvement in the precision of these cosmological parameters.

\section*{Acknowledgments}
This work is supported by the National Natural Science Foundation of China (Grant Nos. 12494575 and U2038106), the Natural Science Foundation of Jiangxi Province of China (grant No. 20242BAB26012), Shandong Provincial Natural Science Foundation (ZR2021MA021) and and Manned Spaced Project (CMS-CSST-2021-A12).

\begin{table*}[htbp]
  \centering
  \caption{The fitting results of reconstructed GRBs.}
    \begin{tabular}{lcccccr}
  \hline
  \hline
    GRB ID   & $z$  & $T_{90}$ & $F_0$ & $t_b$ & $\gamma$ & $\chi^2$\\
     & & (s) & $10^{-11}$(erg/cm$^2$/s) & $10^3$(s) & & \\
  \hline
        050315 & 1.95 & 95.6 & $0.82_{-0.02}^{+0.02}$ & $98.76_{-3.30}^{+3.39}$ & 1.89 & 1.37 \\ \hline
        050319 & 3.24 & 152.5 & $1.61_{-0.08}^{+0.09}$ & $32.58_{-1.64}^{+1.64}$ & 1.85 & 1.28 \\\hline
        050505 & 4.27 & 58.9 & $3.76_{-0.15}^{+0.15}$ & $13.93_{-0.45}^{+0.45}$ & 2.09 & 1.07 \\\hline
        050814 & 5.3 & 150.9 & $0.35_{-0.02}^{+0.02}$ & $41.05_{-2.29}^{+2.48}$ & 1.97 & 1.28 \\\hline
        051008 & 2.77 & $>$32.0 & $7.19_{-0.34}^{+0.37}$ & $5.51_{-0.20}^{+0.21}$ & 1.95 & 1.44 \\\hline
        060526 & 3.22 & 298.2 & $0.83_{-0.04}^{+0.04}$ & $17.49_{-0.86}^{+0.90}$ & 1.89 & 2.27 \\\hline
        060605 & 3.78 & 79.1 & $5.20_{-0.24}^{+0.25}$ & $4.44_{-0.16}^{+0.17}$ & 1.89 & 1.33 \\\hline
        060906 & 3.69 & 43.5 & $0.60_{-0.03}^{+0.03}$ & $14.00_{-0.68}^{+0.71}$ & 2.10 & 2.91 \\\hline
        061222A & 2.09 & 71.4 & $5.10_{-0.21}^{+0.21}$ & $27.82_{-0.88}^{+0.90}$ & 1.84 & 1.33 \\\hline
        070306 & 1.5 & 209.5 & $2.94_{-0.09}^{+0.09}$ & $30.34_{-0.90}^{+0.88}$ & 1.80 & 3.06 \\\hline
        080310 & 2.43 & 365.0 & $1.87_{-0.10}^{+0.11}$ & $13.52_{-0.60}^{+0.63}$ & 2.09 & 1.33 \\\hline
        081008 & 1.97 & 185.5 & $3.63_{-0.26}^{+0.27}$ & $7.38_{-0.41}^{+0.43}$ & 1.98 & 0.88 \\\hline
        090205 & 4.65 & 8.8 & $0.96_{-0.06}^{+0.06}$ & $7.01_{-0.39}^{+0.42}$ & 2.07 & 1.85 \\\hline
        090407 & 1.45 & 310.0 & $0.49_{-0.02}^{+0.02}$ & $56.77_{-2.25}^{+2.31}$ & 2.22 & 1.28 \\\hline
        090516 & 4.11 & 140.0 & $3.95_{-0.17}^{+0.18}$ & $9.50_{-0.34}^{+0.34}$ & 2.03 & 1.32 \\\hline
        100424A & 2.47 & 104.0 & $470.59_{-18.91}^{+18.70}$ & $0.21_{-0.01}^{+0.01}$ & 1.66 & 1.59 \\\hline
        111123A & 3.15 & 290.0 & $0.71_{-0.03}^{+0.03}$ & $18.35_{-0.88}^{+0.91}$ & 2.55 & 1.62 \\\hline
        120404A & 2.88 & 38.7 & $4.89_{-0.32}^{+0.33}$ & $3.24_{-0.16}^{+0.17}$ & 1.90 & 1.20 \\\hline
        130606A & 5.91 & 276.6 & $2.83_{-0.18}^{+0.18}$ & $8.25_{-0.42}^{+0.44}$ & 1.86 & 0.75 \\\hline
        140518A & 4.71 & 60.5 & $3.16_{-0.16}^{+0.17}$ & $3.07_{-0.16}^{+0.17}$ & 2.09 & 2.02 \\\hline
        141026A & 3.35 & 146.0 & $0.27_{-0.01}^{+0.01}$ & $67.29_{-3.05}^{+3.23}$ & 1.92 & 2.84 \\\hline
        151112A & 4.1 & 19.3 & $0.55_{-0.02}^{+0.02}$ & $39.72_{-1.94}^{+1.99}$ & 2.28 & 1.22 \\\hline
        160121A & 1.96 & 12.0 & $0.81_{-0.04}^{+0.04}$ & $13.52_{-0.81}^{+0.86}$ & 2.21 & 3.47 \\\hline
        160227A & 2.38 & 316.5 & $1.32_{-0.05}^{+0.05}$ & $88.37_{-4.07}^{+4.24}$ & 1.67 & 2.25 \\\hline
        161014A & 2.82 & 18.3 & $38.90_{-1.53}^{+1.55}$ & $1.48_{-0.05}^{+0.05}$ & 1.83 & 1.40 \\\hline
        161017A & 2.01 & 216.3 & $5.67_{-0.31}^{+0.33}$ & $10.88_{-0.49}^{+0.51}$ & 1.99 & 1.88 \\\hline
        171222A & 2.41 & 174.8 & $0.09_{-0.01}^{+0.01}$ & $255.03_{-17.90}^{+18.83}$ & 1.99 & 1.19 \\\hline
        180329B & 2.00 & 210.0 & $2.74_{-0.15}^{+0.15}$ & $8.11_{-0.37}^{+0.40}$ & 1.87 & 0.89 \\\hline
        190106A & 1.86 & 76.8 & $4.80_{-0.18}^{+0.18}$ & $29.72_{-1.03}^{+1.06}$ & 1.95 & 2.22 \\\hline
        190114A & 3.38 & 66.6 & $4.09_{-0.19}^{+0.20}$ & $7.03_{-0.32}^{+0.34}$ & 1.83 & 1.29 \\\hline
        190719C & 2.47 & 185.7 & $3.07_{-0.16}^{+0.16}$ & $51.35_{-2.519}^{+2.63}$ & 1.54 & 1.00 \\\hline
        201221A & 5.70 & 44.5 & $2.33_{-0.24}^{+0.26}$ & $7.36_{-0.71}^{+0.78}$ & 1.57 & 1.27 \\\hline
        210210A & 0.72 & 6.6 & $15.11_{-0.56}^{+0.56}$ & $4.06_{-0.16}^{+0.16}$ & 1.82 & 1.25 \\\hline
        210610B & 1.10 & 69.38 & $11.04_{-0.77}^{+0.83}$ & $15.50_{-0.87}^{+0.91}$ & 1.89 & 0.74 \\\hline
        241030A & 1.41 & 173.3 & $21.14_{-2.07}^{+2.30}$ & $6.10_{-0.43}^{+0.46}$ & 1.96 & 2.27 \\\hline
\hline
  \end{tabular}%
  \begin{tablenotes}
  \footnotesize
  \item \textbf{Note.} $z$ represents the redshift of the GRB, and $\gamma$ is the spectral index during the plateau phase, both sourced from the Swift website. $F_0$ is the flux at the end of the plateau, and $t_b$ is the time at which the plateau ends. These two are the best-fit values through the magnetar model.
  \end{tablenotes}
  \label{tab:addlabel}%
\end{table*}%

\begin{table*}[htbp]
  \centering
  \caption{Fitting results of correlation parameters before and after reconstruction.}
    \begin{tabular}{lcccccccr}
  \hline
  \hline
  correlation& $a$  & $b$ & $c$ & $\sigma_{int}$ & $a'$  & $b'$ & $c'$ & $\sigma_{int}'$\\
  \hline
  $L_0-t_b$  & $1.626_{-0.148}^{+0.145}$ & $-0.983_{-0.137}^{+0.137}$ & ~$~$ & $0.380_{-0.063}^{+0.086}$ & $1.594_{-0.155}^{+0.155}$ & $-0.961_{-0.147}^{+0.153}$ & ~$~$ & $0.406_{-0.067}^{+0.093}$\\
  \hline
  $L_0-t_b-E_{p,i}$  & $0.431_{-0.515}^{+0.528}$ & $-0.985_{-0.116}^{+0.118}$ & $0.499_{-0.210}^{+0.203}$ & $0.321_{-0.058}^{+0.079}$ & $0.480_{-0.525}^{+0.516}$ & $-0.975_{-0.125}^{+0.124}$ & $0.489_{-0.211}^{+0.213}$ & $0.332_{-0.057}^{+0.081}$\\
  \hline
  $L_0-t_b-E_{\gamma,\mathrm{iso}}$ & $1.507_{-0.099}^{+0.100}$ & $-1.042_{-0.090}^{+0.093}$ & $0.317_{-0.072}^{+0.073}$ & $0.245_{-0.045}^{+0.059}$ & $1.505_{-0.101}^{+0.102}$ & $-1.031_{-0.099}^{+0.095}$ & $0.326_{-0.077}^{+0.077}$ & $0.259_{-0.043}^{+0.063}$\\
  \hline
    \end{tabular}%
      \begin{tablenotes}
  \footnotesize
  \item \textbf{Note.} $a$ ($a'$) represents the best-fit value before (after) reconstruction.
        \end{tablenotes}
  \label{tab:addlabe2}%
\end{table*}%

\begin{table*}[htbp]
  \centering
  \caption{Constraints on cosmological parameters before and after reconstruction.}
    \begin{tabular}{lcccc}
  \hline
  \hline
  correlation& $\Omega_m$  & $\Omega_m/\Omega_\Lambda$ & $\Omega_m'$  & $\Omega_m'/\Omega_\Lambda'$\\
  \hline
  $L_0-t_b$  & $0.290_{-0.008}^{+0.008}$ &  $0.331_{-0.033}^{+0.033}$ / $0.762_{-0.041}^{+0.040}$  &  $0.290_{-0.008}^{+0.008}$  & $0.333_{-0.033}^{+0.033}$ / $0.763_{-0.042}^{+0.041}$ \\
  \hline
  $L_0-t_b-E_{p,i}$  & $0.290_{-0.008}^{+0.008}$ &  $0.335_{-0.035}^{+0.036}$ / $0.767_{-0.044}^{+0.044}$  &  $0.290_{-0.008}^{+0.007}$  & $0.334_{-0.035}^{+0.036}$ / $0.766_{-0.043}^{+0.044}$ \\
  \hline
  $L_0-t_b-E_{\gamma,\mathrm{iso}}$ & $0.290_{-0.007}^{+0.007}$ &  $0.328_{-0.030}^{+0.030}$ / $0.758_{-0.037}^{+0.037}$  & $0.290_{-0.008}^{+0.007}$  & $0.329_{-0.030}^{+0.032}$ / $0.759_{-0.039}^{+0.038}$\\
  \hline
    \end{tabular}%
      \begin{tablenotes}
  \footnotesize
  \item \textbf{Note.} The first column represents different correlations, the second column represents the constraint results of cosmological parameters under a flat $\Lambda$CDM model, the third column represents the constraint results of cosmological parameters under a non-flat $\Lambda$CDM model, and the fourth and fifth columns represent the constraint results of cosmological parameters after reconstruction under flat and non-flat $\Lambda$CDM models, respectively.
        \end{tablenotes}
  \label{tab:addlabe3}%
\end{table*}%

\begin{table*}[htbp]
  \centering
  \caption{Changes in uncertainty after reconstruction.}
    \begin{tabular}{lcccccc}
  \hline
  \hline
  correlation& $EF_{\Omega_m}$  & $EF_{\Omega_m}/EF_{\Omega_\Lambda}$ & $EF_{\Omega_m}'$  & $EF_{\Omega_m}'/EF_{\Omega_\Lambda}'$ &  $\%_{\Omega_m} $ & $\%_{\Omega_m}/\%_{\Omega_\Lambda}$\\
  \hline
  $L_0-t_b$  &  0.028 & 0.100/0.053 & 0.028 & 0.099/0.054 & 0 & -0.6/2.33 \\ \hline
  $L_0-t_b-E_{p,i}$  & 0.028 & 0.106/0.057 & 0.026 & 0.106/0.057 & -6.25 & 0.30/-1.01 \\ \hline
  $L_0-t_b-E_{\gamma,\mathrm{iso}}$ & 0.024 & 0.091/0.049 & 0.026 & 0.097/0.051 & 7.14 & 6.34/3.92 \\ \hline
    \end{tabular}%
      \begin{tablenotes}
  \footnotesize
  \item \textbf{Note.} $EF_{\Omega_m}$ ($EF_{\Omega_m}'$) represents the error fraction of $\Omega_m$ before (after) reconstruction, and $\%_{\Omega_m}$ represents the change in uncertainty of $\Omega_m$.
        \end{tablenotes}
  \label{tab:addlabe4}%
\end{table*}%

\end{document}